\documentclass{aa}
\usepackage[]{natbib}
\usepackage{graphicx,color}
\usepackage{txfonts}
\usepackage[colorlinks=true, citecolor=blue]{hyperref}

\begin{document} 

\title{\texttt{Jurassic}: A Chemically Anomalous Structure in the Galactic Halo} 

	\author{
		Jos\'e G. Fern\'andez-Trincado\inst{1, 2}\thanks{To whom correspondence should be addressed; E-mail: jfernandez@obs-besancon.fr and/or jose.fernandez@uda.cl},
		Timothy C. Beers\inst{3}
		and 
		Dante Minniti\inst{4,5}
	}
	
	\authorrunning{J. G. Fern\'andez-Trincado et al.} 
	
\institute{
       	Institut Utinam, CNRS-UMR 6213, Universit\'e Bourgogne-Franche-Compt\'e, OSU THETA Franche-Compt\'e, Observatoire de Besan\c{c}on, BP 1615, 251010 Besan\c{c}on Cedex, France
       	\and
		Instituto de Astronom\'ia y Ciencias Planetarias, Universidad de Atacama, Copayapu 485, Copiap\'o, Chile
		\and
	    Department of Physics and JINA Center for the Evolution of the Elements, University of Notre Dame, Notre Dame, IN 46556, USA
	    \and
	    Depto. de Cs. F\'isicas, Facultad de Ciencias Exactas, Universidad Andr\'es Bello, Av. Fern\'andez Concha 700, Las Condes, Santiago, Chile
	    \and
	    Vatican Observatory, V00120 Vatican City State, Italy
    }
	
	\date{Received ...; Accepted ...}
	\titlerunning{Elemental Abundances of an Accreted Structure}
	
	
	\abstract
	{Detailed elemental-abundance patterns of giant stars in the Galactic halo measured by APOGEE-2 have revealed the existence of a unique and significant stellar sub-population of silicon-enhanced ([Si/Fe]$\gtrsim +0.5$) metal-poor stars, spanning a wide range of metallicities ($-1.5\lesssim$[Fe/H]$\lesssim-0.8$). Stars with over-abundances in [Si/Fe] are of great interest because these have very strong silicon ($^{28}$Si) spectral features for stars of their metallicity and evolutionary stage, offering clues about rare nucleosynthetic pathways in globular clusters (GCs). Si-rich field stars have been conjectured to have been evaporated from GCs, however, the origin of their abundances remains unclear, and several scenarios have been offered to explain the anomalous abundance ratios. These include the hypothesis that some of them were born from a cloud of gas previously polluted by a progenitor that underwent a specific and peculiar nucleosynthesis event, or due to mass transfer from a previous evolved companion. However, those scenarios do not simultaneously explain the wide gamut of chemical species that are found in Si-rich stars. Instead, we show that the present inventory of such unusual stars, as well as their relation to known halo substructures (including the in-situ halo, \textit{Gaia}-Enceladus, the Helmi Stream(s), and Sequoia, among others) is still incomplete. We report the chemical abundances of the iron-peak (Fe), the light- (C and N), the $\alpha-$ (O and Mg), the odd-Z  (Na and Al), and the \textit{s}-process (Ce and Nd) elements of 55 newly identified Si-rich field stars (among more than $\sim$600,000 APOGEE-2 targets), that exhibit over-abundances of [Si/Fe] as extreme as those observed in some Galactic GCs, and are relatively cleanly from other stars in the [Si/Fe]-[Fe/H] plane. This new census confirms the presence of a statistically significant and chemically-anomalous structure in the inner halo: \texttt{Jurassic}. The chemo-dynamical properties of the \texttt{Jurassic} structure is consistent with it being the tidally disrupted remains of GCs, easily distinguished by an over-abundance of [Si/Fe] among Milky Way populations or satellites.
	}
	
	\maketitle
	
	\section{Introduction}
	\label{section1}
	
	The stellar content of the halo of the Milky Way (MW) is littered by a mixture of stellar debris of completely and/or partially destroyed dwarf galaxies and GCs \citep[e.g.,][]{Helmi1999, Carollo2007, Carollo2010, Nissen2010, Fernandez-Trincado2013, Fernandez-Trincado2015a, Fernandez-Trincado2015b, Fernandez-Trincado2016a, Fernandez-Trincado2016b, Fernandez-Trincado2017a, Recio-Blanco2017, Bekki2019, Fernandez-Trincado2019b, Fernandez-Trincado2019c, Fernandez-Trincado2019d, Koch2019, Massari2019, Hanke2020, Thomas2020, Yuan2020b, Wan2020, Naidu2020, Fernandez-Trincado2020a, Fernandez-Trincado2020b, Fernandez-Trincado2020c}, which preserve signatures of the Galaxy's assembly history \citep[see][for a recent review]{Naidu2020}. 
	
	Most of the stellar halo debris identified to date extend out to several hundred kiloparsecs, and have been identified in a wide variety of discrete structures, preserving important insight into the earliest accretion events. \textit{Gaia} Data Release 2 \citep[DR2;][]{Brown2018}, complemented with ground-based spectroscopic surveys, has provided compelling evidence that both the inner- and outer-halo populations of the MW are dominated by an \textit{ex-situ} formation scenario--being built up through massive ($>10^{10}$ M$_{\odot}$) accretion events that have probably occurred $\sim$8--11 Gyr ago, including the \textit{Gaia}-Enceladus (G-E) dwarf galaxy \citep{Belokurov2018, Haywood2018, Helmi2018, Koppelman2018, Myeong2018}, accompanied by other significant merger events such as the \texttt{Sequoia} \citep{Myeong2019}, the \texttt{Helmi} stream(s) \citep{Helmi1999, Chiba2000, Koppelman2019b}, the \texttt{Kraken} \citep{Kruijssen2019}, \texttt{Thamnos} 1 \& 2 \citep{Koppelman2019}, the disrupting \texttt{Sagittarius} dwarf galaxy \citep{Ibata1994, Law2010, Majewski2013, Hasselquist2019, Hayes2020}, as well as streams recently identified as debris from very metal-poor globular clusters \citep{Thomas2020, Wan2020, Yuan2020b}, and the \texttt{Fimbulthul} structure associated with the unusual globular cluster $\omega$ Cen \citep{Ibata2019}. In this paper, we add to this collection a chemically anomalous halo population of Si-rich stars \citep{Fernandez-Trincado2019d}, which we  refer to as the \texttt{Jurassic} structure. 
	
	There also exists a wealth of observational evidence for a possible "\textit{in-situ}" channel, especially for the inner-halo population itself, which is spatially, kinematically, and chemically distinguishable from the outer-halo population \citep{Carollo2007, Carollo2010, Beers2012, An2013, An2015, An2020}, and thought to have formed in part from gas accreted by the MW at early times \citep{Carollo2013, Tissera2014, Hawkins2015, Hayes2018, Alvar2018, Alvar2019}. These studies illustrate the complex formation history of the stellar halo of the MW, which may involve a mixture of stars likely formed \textit{in-situ} and stellar debris accreted from different structures. 
	
	Although the morphology and chemo-dynamical properties of the stellar halo have been extensively explored over the past few decades (see \citealt{Beers2005}, \citealt{Ivezic2012}, and \citealt{Helmi2020} for reviews), a long-standing problem remains as to how the different formation scenarios for the stellar halo may be discriminated from one another.  Current large stellar spectroscopic surveys of the MW, such as the Apache Point Observatory Galactic Evolution Experiment \citep[APOGEE:][]{Majewski2017}, enable methods such as chemical tagging, which is based on the principle that the chemical composition of stars reflect the site of their formation \citep{Freeman2002, Feltzing2013, Ting2015, Hogg2016}, can help clarify the picture of where the majority of halo stars are produced, and what caused them to achieve their current physical properties. 
	
	The search for field stars that were born in GCs \citep{Lind2015, Martell2016, Fernandez-Trincado2015b, Fernandez-Trincado2016a, Fernandez-Trincado2016b, Fernandez-Trincado2017a, Fernandez-Trincado2019a, Fernandez-Trincado2019b, Fernandez-Trincado2019c, Fernandez-Trincado2019d, Fernandez-Trincado2020a, Simpson2020} is one clear example of the power of chemical tagging. This is possible because GCs appear to be the only environment responsible for the presence of light-element anti-correlations at all stellar evolutionary phases \citep[e.g.,][]{Martell2016, Pancino2017, Bastian2018, Masseron2019, Szabolcs2020}, unless they are part of a binary system \citep{Bastian2018, Fernandez-Trincado2019b}, or part of the new kind of recently discovered anomalous Phosphorus-rich field stars \citep[see, e.g.,][]{Masseron2020}. Thus, a complete census of all those chemically anomalous stars will help develop a better understanding for the assembly of the inner and outer halo, where substantial amount of stellar debris from GCs are thought to currently reside \citep[see, e.g.,][]{Martell2016, Fernandez-Trincado2019c, Fernandez-Trincado2019d, Fernandez-Trincado2020a}.

	In this paper, we update the census of silicon-enriched metal-poor stars, making use of data from the APOGEE-2 survey \citep{Majewski2017}. The silicon-enriched metal-poor stars are of particular interest as they belong to the exclusive collection of metal-poor stars in GCs where $^{28}$Si leaking from the Mg-Al cycle at high temperature \citep[see e.g.][]{Meszaros2015, Szabolcs2020}. Thus, such a search in other Galactic environments provides strong evidence either for or against the uniqueness of the progenitor stars to GCs evolution. This paper is outlined as follows. In Section \ref{section2}, we present details and information regarding the APOGEE-2 data set and the incremental data used in this work. In Section \ref{section3}, we describe our sample selection, the adopted stellar parameters, and the methodology employed to determine the elemental abundances. In Section \ref{section4}, we present the newly identified Si-rich stars. In Section \ref{section5}, we present the statistical significance of the \texttt{Jurassic} structure. In Section \ref{section6}, we present details regarding the elemental abundances of the stars in the \texttt{Jurassic} structure. In Section \ref{section7}, we present a dynamical study of the stars in  the \texttt{Jurassic} structure. Our conclusions are drawn in Section \ref{section8}.

	\section{The APOGEE \textit{H}-band Spectroscopic survey}
	\label{section2}

	The dataset analysed in this work comes primarly from the Apache Point Observatory Galactic Evolution Experiment \citep[APOGEE:][]{Zasowski2013, Majewski2017} and its successor APOGEE-2 \citep{Zasowski2017}, which have collected high-resolution ($R\sim$22,500) H-band spectra (near-IR, $\sim$15,145 $\AA{}$ to 16,960 $\AA{}$, vacuum wavelengths) for almost 430,000 sources in their sixteenth data release \citep[DR16,][]{Ahumada2020}, as part of the Sloan Digital Sky Survey IV \citep[][]{Blanton2017}. Here we take advantage of new data taken subsequent to the DR16 release, which were reduced with the same pipeline as the DR16 stars. This new incremental data set provides to the scientific community spectra of more than 680,000 stars; we refer to these data as the incremental APOGEE-2 DR16 plus (hereafter APOGEE-$2+$). 
	
	APOGEE-$2+$ includes data taken from both the Northern and Southern hemisphere using the APOGEE-$2$ spectrographs \citep{Eisenstein2011, Wilson2012, Wilson2019} mounted on the 2.5m Sloan Foundation telescope \citep{Gunn2006} at Apache Point Observatory in New Mexico (APOGEE-2N: North, APO), and in the 2.5m Ir\'en\'ee du Pont telescope \citep{Bowen1973} at Las Campanas Observatory (APOGEE-2S: South, LCO) in Chile. For details regarding the APOGEE atmospheric-parameter analysis we direct the reader to the description of the APOGEE Stellar Parameter and Chemical Abundances  pipeline \citep[\texttt{ASPCAP}:][]{GarciaPerez2016a}, while for details about the grid of synthetic spectra and errors see \citet{Holtzman2015}, \citet{Holtzman2018}, \citet{Henrik2018}, and \citet{Henrik2020}. We also refer the reader to \citet{Nidever2015} for further details regarding the data reduction pipeline for APOGEE$2+$. The model grids for APOGEE-$2+$ are based on a complete set of \texttt{MARCS} \citep{Gustafsson2008} stellar atmospheres, which now extend to effective temperatures as low as 3200 K, and spectral synthesis using the \texttt{Turbospectrum} code \citep{Plez2012}.
	
	\section{Data} 
	\label{section3}

	Since we are primarily interested in the detection and mapping of Si-rich metal-poor stars, our focus in this work is on giants in the metallicity range between [Fe/H] = $-$1.8 and $-$0.7. The Si-rich stars were first discovered in \citet[][]{Fernandez-Trincado2019d}, and were hypothesized to belong to a new sub-population of the inner stellar halo. Here, we conduct a large search for such stars in the APOGEE-$2+$ catalogue. By imposing a lower limit on metallicity, [Fe/H] $>$ $-$1.8, we include stars with high-quality spectra and reliable parameters and abundances. The requirement of an upper limit of [Fe/H] $<$ $-$0.7 minimizes the presence of stars belonging to the disk system. Thus, we selected a sample of giant stars, adopting conservative cuts on the columns of the APOGEE-$2+$ catalogue in the following way: 
	
	\begin{itemize}
		\item[(i)] S/N $>$ 60 pixel$^{-1}$. This cut was chosen to ensure that we are selecting spectra that have well-known uncertainties in their stellar parameters and chemical abundances, and remove stars with lower quality spectra \citep[e.g.,][]{GarciaPerez2016a}. 
		\item[(ii)] 3200  K $ < T^{ASPCAP}_{\rm eff} < $  6000 K. This temperature range ensures that the stellar parameters are reliably and consistently determined, and maximizes the overall quality of the abundances considered \citep{GarciaPerez2016a, Holtzman2018}. 
		\item[(iii)] The estimated $\log$ \textit{g}$^{ASPCAP}$ must be less than 3.6. This cut was chosen to ensure that stars have more accurate \texttt{ASPCAP}-derived parameters than the stars with $\log$ \textit{g}$^{ASPCAP}$ $>$ 3.6.  Due to the lack of asteroseismic surface gravities for dwarfs, only stars with  $\log$ \textit{g} $<$ 3.6 have calibrated surface gravity estimates \citep[see e.g.,][]{Henrik2018, Henrik2020}.
		\item[(iv)] \texttt{ASPCAPFLAG} $==$ \texttt{0}. This cut ensures that there were no major flagged issues, i.e., low signal-to-noise, poor synthetic spectral fit, stellar parameters near grid boundaries, among others \citep[e.g.,][]{Holtzman2015, GarciaPerez2016a}.
		\item[(v)] For our selection, we also disregarded stars in GCs from our sample, i.e., those sources analyzed in \citet[][]{Masseron2019} and \citet{Meszaros2020}.
		\item[(vi)] Lastly, field stars identified previously as P-, N- and Si-rich stars from \citet{Martell2016}, \citet{ Fernandez-Trincado2016b}, \citet{Fernandez-Trincado2017a, Fernandez-Trincado2019a, Fernandez-Trincado2019b, Fernandez-Trincado2019c, Fernandez-Trincado2019d}, and \citet{Masseron2020} were excluded from our sample.
	\end{itemize}

Our initial sample contains about 19,700 stars,  which is four time larger than the sample of \citet{Fernandez-Trincado2019d} analyzed in previous data releases. After the internal release of the APOGEE-$2+$ catalogue, we discovered many more stars belonging to the Si-rich sample, which we report here as part of a larger homogeneous census. We have added these new stars to the sample, derived new atmospheric parameters, and ran our abundance determination pipeline \texttt{BACCHUS} \citep{Masseron2016} with the same setup used in \citet{Fernandez-Trincado2019d}. Here, we proceed with a detailed re-analysis of the newly discovered Si-rich stars with the \texttt{BACCHUS} code, as \texttt{ASPCAP} introduce its own set of problems for the effective temperatures and [X/Fe] at low metallicities, [Fe/H]$< -0.7$ dex \citep[see, e.g.][]{Fernandez-Trincado2019b, Fernandez-Trincado2019c, Fernandez-Trincado2019d, Nataf2019, Meszaros2015, Szabolcs2020, Fernandez-Trincado2020b}.

\subsection{Stellar Parameters and Abundance Determinations} 
\label{abundances}
	
Since the method of deriving atmospheric parameters and abundances is identical to that as described in \citet{Fernandez-Trincado2019d}, we only provide a short overview of it in this paper. As before \citep{Fernandez-Trincado2016b, Fernandez-Trincado2017a, Fernandez-Trincado2019a, Fernandez-Trincado2019b, Fernandez-Trincado2019c, Fernandez-Trincado2019d}, we use the uncalibrated stellar parameters for $T^{ASPCAP}_{\rm eff}$ (\texttt{FPARAM\_1}) and $\log{}$ \textit{g}$^{ASPCAP}$ (\texttt{FPARAM\_2}). First, we made a careful inspection of each \textit{H}-band spectrum with the \texttt{BACCHUS} code to derive the metallicity, broadening parameters, and chemical abundances, based on a line-by-line approach in the same manner as in \citet{Fernandez-Trincado2019d}, and summarized here for guidance. The \texttt{BACCHUS} code relies on the radiative transfer code \texttt{Turbospectrum} \citep{Alvarez1998, Plez2012} and the \texttt{MARCS} model atmosphere grid \citep{Gustafsson2008}. 

For each element and each line, the abundance determination proceeds as in \citet{Hawkins2016} and \citet{Fernandez-Trincado2019c}: (\textit{a}) A spectrum synthesis, using the full set of atomic and molecular line lists described in \citet{Shetrone2015}, (Neodymium: Nd II) \citet[][]{Hasselquist2016} and (Cerium: Ce II) \citet[][]{Cunha2017} (this set of lists is internally labeled as \texttt{linelist.20170418} based on the date of creation in the format YYYYMMDD). This is used to find the local continuum level via a linear fit; (\textit{b}) Cosmic and telluric rejections are performed; (\textit{c}) The local S/N is estimated; (\textit{d}) A series of flux points contributing to a given absorption line is automatically selected; and (\textit{e}) Abundances are then derived by comparing the observed spectrum with a set of convolved synthetic spectra characterised by different abundances. Then, four different abundance determination methods are used: (\textit{1}) line-profile fitting; (\textit{2}) core line-intensity comparison; (\textit{3}) global goodness-of-fit estimate; and (\textit{4}) equivalent-width comparison. Each diagnostic yields validation flags. Based on these flags, a decision tree then rejects or accepts each estimate, keeping the best-fit abundance. Here, we adopted the $\chi^2$ diagnostic as the abundance because it is the most robust. However, we store the information from the other diagnostics, including the standard deviation between all four methods.
	
A mix of heavily CN-cycled and $\alpha$-poor \texttt{MARCS} models were used, as well as the same molecular lines adopted by \citep{Smith2013}, and employed to determine the C, N, and O abundances.  In addition, we have adopted the C, N, and O abundances that satisfy the fitting of all molecular lines consistently; i.e., we first derive $^{16}$O abundances from $^{16}$OH lines, then derive $^{12}$C from $^{12}$C$^{16}$O lines, and $^{14}$N from $^{12}$C$^{14}$N lines; the CNO abundances are derived several times to minimize the OH, CO, and CN dependences \citep[see, e.g.,][]{Smith2013, Fernandez-Trincado2016b, Fernandez-Trincado2017a, Fernandez-Trincado2019a, Fernandez-Trincado2019b, Fernandez-Trincado2019c, Fernandez-Trincado2019d, Fernandez-Trincado2020d}.
	
Lastly, to provide a consistent chemical analysis, we re-determine the chemical abundances, assuming as input the uncalibrated effective temperature ($T^{ASPCAP}_{\rm eff}$ or \texttt{FPARAM\_1}), surface gravity ($\log$ \textit{g}$^{ASPCAP}$ or \texttt{FPARAM\_2}), and metallicity ([M/H] or \texttt{FPARAM\_3}) as derived by \texttt{ASPCAP}/APOGEE-$2+$ run. 

We also applied a simple approach of fixing $T^{pho}_{\rm eff}$ and $\log$ \textit{g} to values determined independently of spectroscopy, in order to check for any significant deviation in the chemical abundances. For this, the photometric effective temperatures were calculated from the $J_{2MASS}-K_{s,2MASS}$ colour relation using the methodology presented in \citet{Gonzalez2009}. 

Photometry is extinction corrected using the Rayleigh Jeans Color Excess (RJCE) method \citep{Majewski2011}. We estimate surface gravity from 10 Gyr \texttt{PARSEC} \citep{Bressan2012} isochrones, as 10 Gyr is the typical age of Galactic GCs \citep[see, e.g.,][]{Baumgardt2019}; here we assume that the stars in the \texttt{Jurassic} structure could be possible GC migrants. 

The adopted stellar parameters are listed in Table \ref{Table1}. It is also important to note that the absence of radial-velocity variation, as listed in the same table (visit-to-visit variation, $RV_{SCATTER} < 2$ km s$^{-1}$), does not support any evidence for a binary companion, i.e., none of the newly identified Si-rich giants  in the \texttt{Jurassic} structure  has a strong variability in its radial velocity over the period of the APOGEE-$2+$ observations ($\lesssim6$ months). However, long-term radial-velocity monitoring of all of our stars would naturally be the best course to establish the number of such objects formed through the binary channel. 
	
\begin{figure}
	\begin{center}
		\includegraphics[width=100mm]{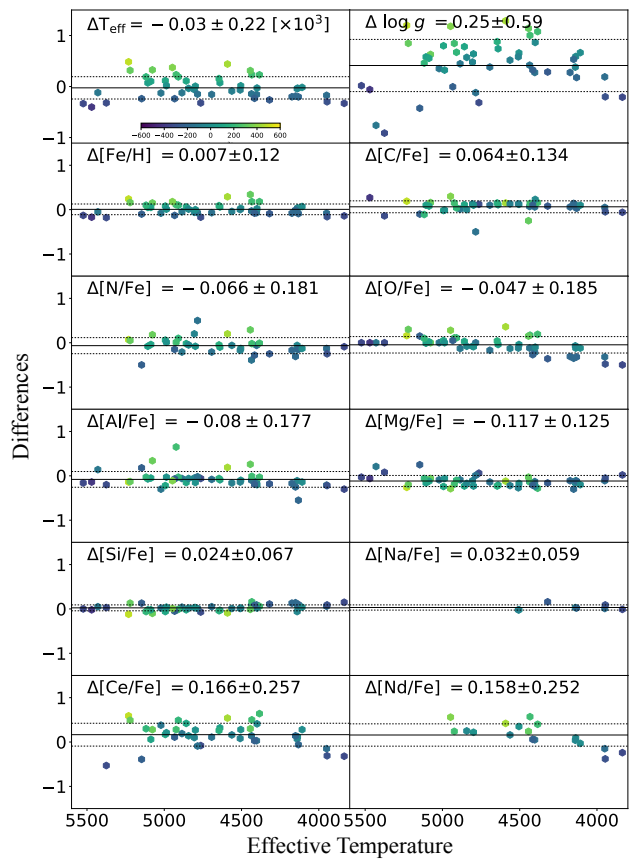}
		\caption{The newly identified stars in the \texttt{Jurassic} structure analyzed in this work. Differences in atmospheric parameters and elemental abundances produced by two runs adopting different effective temperatures ($T_{\rm eff}$) and surface gravities ($\log$ \textit{g}): photometric versus uncalibrated \texttt{ASPCAP} values. The symbols are colour-coded by the differences between the photometric and uncalibrated \texttt{ASPCAP} temperatures. The average and $\pm$ errors (the standard deviation around the mean of the differences) are listed in the title of each panel.}
		\label{Figure1}
	\end{center}
\end{figure}

Figure \ref{Figure1} compares the sensitivity to the derived atmospheric parameters, depending on the species and line in question. When the spectroscopic and photometry-based atmospheric parameters were adopted, we found large discrepancies in the effective temperature ($\gtrsim200$ K) and surface gravity ($\gtrsim0.6$ dex) for a few stars, particularly for hotter ($\gtrsim5000$ K) stars. This issue does not strongly affect the derived [Si/Fe] abundance ratios, but other chemical species such as nitrogen, oxygen, aluminum, cerium, and neodymium are more ($\gtrsim0.2$ dex) affected by these atmospheric discrepancies. However, as can be seen in the same figure, these large discrepancies do not have a strong impact on our determined [Si/Fe] abundance ratios, whose differences are $<0.07$ dex, much less that the reported intrinsic error. 

Figure \ref{Figure3} confirms the reliability in the detected Si I lines, where the spectra of some Si-rich and Si-normal stars are compared in the relevant wavelength intervals. The Si-rich stars have remarkably stronger Si l lines which, in view of the similarity between the pairs of stars in all the other relevant parameters, can only mean that they have much higher silicon abundances. 

\begin{figure*}
	\begin{center}
		\includegraphics[width=210mm]{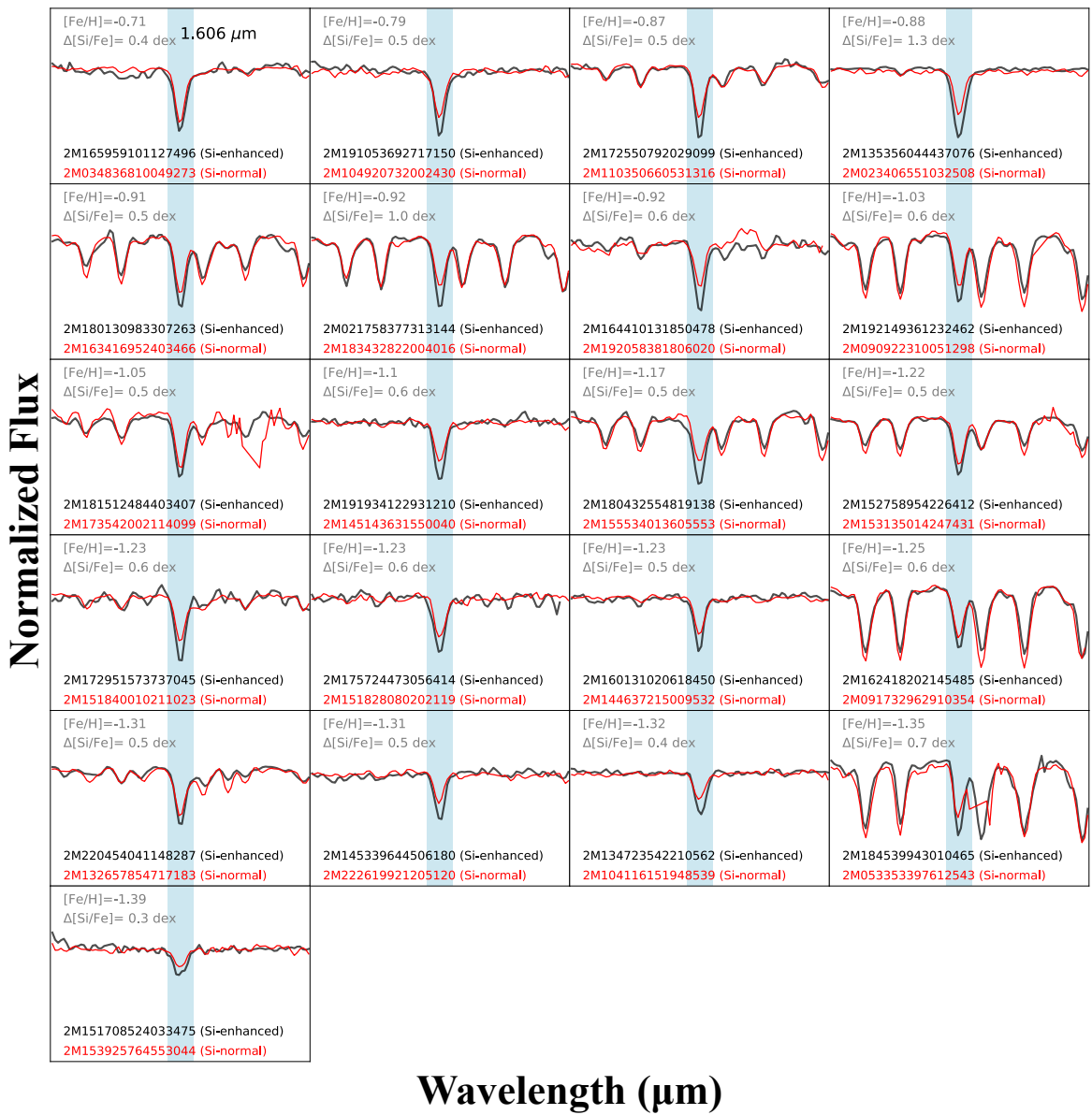}
		\caption{Portions of 21 Si-rich stars (black) compared to Si-normal (red) stars, with similar stellar parameters, around the Si I absorption line at 1.606 $\mu$m (blue shadow region). The stars are ordered by decreasing metallicity, which is indicated in the top left corner in each panel. The $\Delta$[Si/Fe] between both stars is also shown.}
		\label{Figure2}
	\end{center}
\end{figure*}

Figure \ref{Figure1} also indicates that the \texttt{BACCHUS} code recovers the [Fe/H] abundance ratios of these stars within $\sim$0.12 dex for both the adopted photometric and spectroscopic temperatures. The adoption of a purely photometric temperature scale enables us to be somewhat independent of the \texttt{ASPCAP} pipeline, which provides important comparison data for future pipeline validation. However, as pointed out by \citet{Meszaros2020}, the use of photometric temperatures introduces their own set of potential problems related to high E(B$-$V) values, as the \citet{Gonzalez2009} relations are very sensitive to small changes in E(B$-$V). 

Table \ref{Table1} shows that $\sim$20\% of the stars in the \texttt{Jurassic} structure have E(B$-$V) $>$ 0.4, so either the reddening and/or the photometric temperatures are not reliable. For this reason, we limit our discussion to stars in the \texttt{Jurassic} structure with abundance determinations from spectroscopic atmospheric parameters, similar to our previous papers. The complete set of abundances for ten chemical species -- C, N, O, Na, Mg, Al, Si, Fe, Ce, and Nd -- can be found in Table \ref{Table2}. 
	
	 Tables \ref{Table3} and \ref{Table4} list an example of the typical uncertainties for twenty six randomly selected stars in our sample, defined as:
		
	\begin{equation}
		\sigma^{2}_{total}  = \sigma^2_{[X/H], T_{\rm eff}}    + \sigma^2_{[X/H],{\rm log} g} + \sigma^2_{[X/H],\xi_t}  + \sigma^2_{mean}  
	\end{equation}
	
	\noindent
where $\sigma^2_{mean}$ is calculated using the standard deviation derived from the different abundances of the different lines for each element. The values of $\sigma^2_{[X/H], T_{\rm eff}}$, $\sigma^2_{[X/H],{\rm log} g}$, and $\sigma^2_{[X/H],\xi_t} $ are derived for the elements in each star using the sensitivity values of $\pm100$ K for the temperature, $\pm0.3$ dex for log \textit{g}, and 0.05 km s$^{-1}$ for the microturbulent velocity ($\xi_{t}$). 

It is important to note that our results are compared with [X/Fe] abundance ratios determined with the \texttt{ASPCAP}, thus, in order to proceed with an appropriate comparison we correct the \texttt{ASPCAP} by the typical offset of each chemical species between the \texttt{BACCHUS} and \texttt{ASPCAP} pipeline found for a control sample of $\sim$1000 metal-poor ($-1.8\lesssim$[Fe/H]$\lesssim-0.7$) stars belonging to the main components of the MW (halo, disk, and bulge). At the same manner as in \citet{Fernandez-Trincado2020b}, we find that \texttt{ASPCAP} significantly underestimates most of the chemical species by about $\sim$0.1 to 0.3 dex for most of the metal-poor stars \citep[see also][]{Nataf2019}. Such offsets were taken into consideration for the whole MW stars from \texttt{ASPCAP} determinations.

\section{Stars in the \texttt{Jurassic} Structure} 
	\label{section4}
	
\begin{figure}	
	\begin{center}
		\includegraphics[width=95mm]{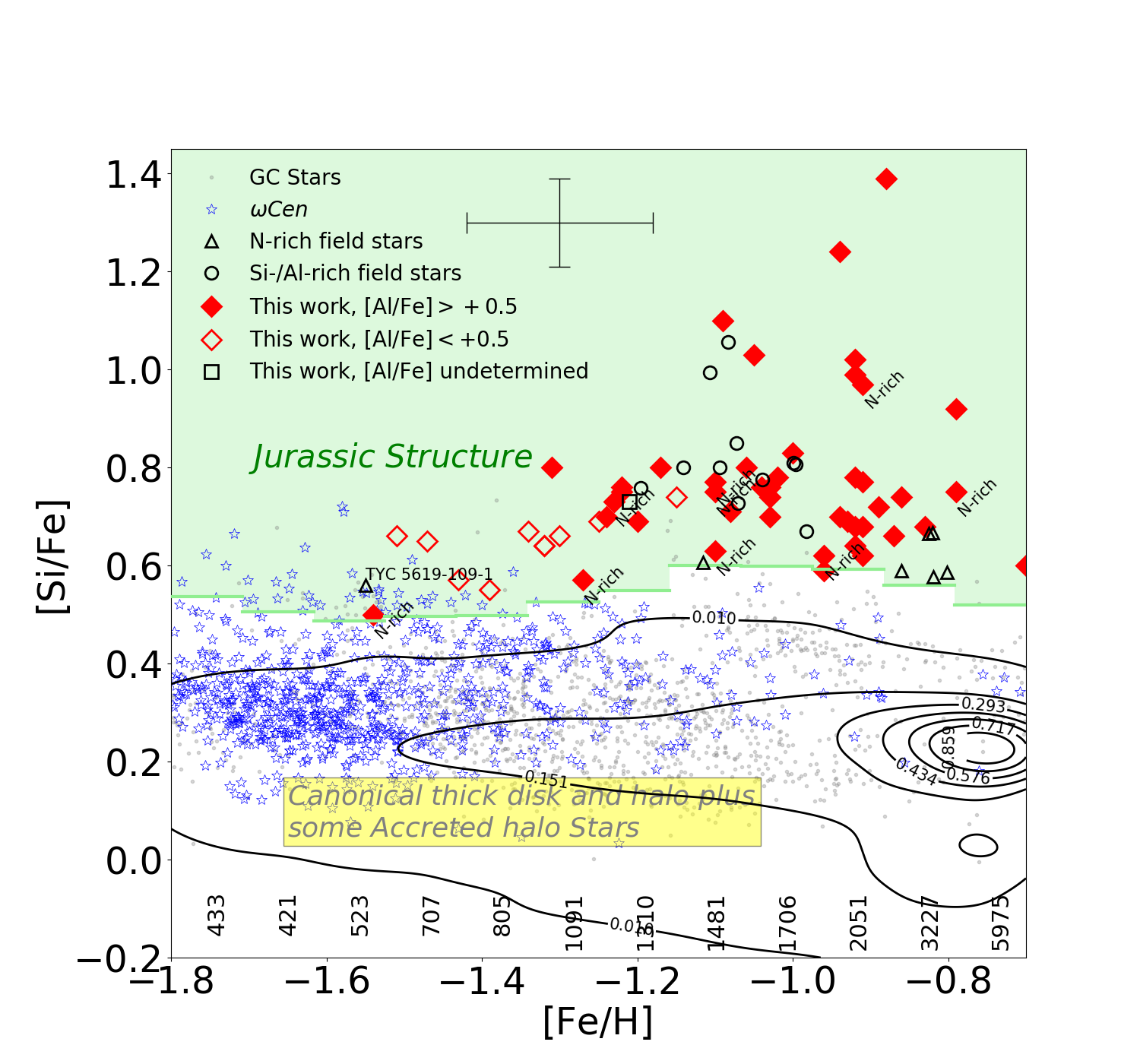}
		\caption{Kernel Density Estimation (KDE) of the \texttt{ASPCAP}-derived [Si/Fe] abundances for MW stars, as a function of [Fe/H], with contours showing the density of stars that appear to be a metal-poor extension of the thick disk and halo locus. These contours demonstrate that there is a low-density boundary (light green thick lines) separating two higher density regions, one with lower [Si/Fe] (black contours), and one with higher [Si/Fe] (light green shadow region). The \texttt{BACCHUS}-derived [Si/Fe] abundance for the newly identified Si-enhanced stars is highlighted with a black empty square (undetermined [Al/Fe]), red empty diamonds ([Al/Fe]$\lesssim+0.5$ in both the intermediate- and low-aluminum regimes, which are likely to be chemically tagged as escaped former members of MW dwarf spheroidal satellites or GCs), and red filled diamonds (with [Al/Fe]$\gtrsim+0.5$ or Al-enhanced stars, which are likely to be chemically tagged as migrants from GCs). Over-plotted are data from \citet{Fernandez-Trincado2016b, Fernandez-Trincado2017a, Fernandez-Trincado2019c} (empty black triangles), from \citet{Fernandez-Trincado2019d} (empty black circles), and GC stars from \citet{Meszaros2020} (grey dots and empty blue star symbols--highlighting the $\omega$ Centauri population). The plotted error bars show the median abundance uncertainty from \texttt{BACCHUS}. The number of APOGEE-2$+$ stars each bin are shown at the center of the bin at the bottom of the plot.}
		\label{Figure3}
	\end{center}
\end{figure}	

Figure \ref{Figure3} shows [Si/Fe] versus [Fe/H], as derived from the \texttt{ASPCAP} pipeline, for our initial sample. The overall behaviour for [Si/Fe] as a function of [Fe/H] shows a smooth transition between the thick-disk and halo population with low-[Si/Fe] ($\lesssim+0.5$) abundance ratios, which we refer to as Si-normal stars in the main MW. 

Our search for Si-rich stars begins with a silicon- and metallicity-based selection criterion in the same manner as in \citet{Fernandez-Trincado2019d}. Using over $\sim$19,700 stars, we determined the boundary between the Si-rich and Si-normal field stars by identifying the trough in the [Si/Fe] distribution in twelve metallicity bins. Figure \ref{Figure3} shows the boundary (light green thick lines) and the number of stars in each of the metallicity bins used to determine the separation between the Si-rich and the Si-normal sequences. The boundary between these two sequences was determined by estimating the average and the standard deviation in [Si/Fe] per metallicity bin in the main body, i.e., the light green thick lines in Figure \ref{Figure3} should be understood as $\langle$[Si/Fe]$\rangle$$_{\rm bin}$ $+$ 3$\sigma_{\rm [Si/Fe]}$ per metallicity bin. The bin sizes were chosen to ensure at least 400 stars were in each bin, with bin centres at [Fe/H]$=$ $-$1.75 to $-$0.75, in steps of 0.09 dex (the number of stars per metallicity bin are shown at the bottom in the same figure). We then label all stars with silicon over-abundances more than $+$3$\sigma_{\rm [Si/Fe]}$ above the $\langle$[Si/Fe]$\rangle$$_{\rm bin}$ at fixed metallicity as Si-rich, which is the same as the selection in \citet{Fernandez-Trincado2019d}. This returns 55 stars as potential Si-rich stars relative to the final data set, which we have called the \texttt{Jurassic} structure. 

Our first selection was simply based on \texttt{ASPCAP} abundances, however, for several other issues that might affect the abundance determinations in the metal-poor regime \citep[see e.g.,][]{Henrik2018}, we have decided to adopt a detailed manual examination and visual inspection, in order to ensure that the spectral fit was adequate for those stars by using the Brussels Automatic Stellar Parameter (\texttt{BACCHUS}) code \citep{Masseron2016} to re-derive the chemical abundances of our sample, by adopting a simple line-by-line approach of selected atomic and molecule lines, under the assumption of local thermodynamic equilibrium (LTE), and using the standard iron ionization -- excitation equilibrium technique. If the lines were not well-reproduced by the synthesis, or the lines were strongly blended or too weak in the spectra of stars to deliver reliable [Si/Fe] ratios, they were rejected. Thus, the [X/Fe] abundance ratios relies on abundance determinations from the \texttt{BACCHUS} pipeline, and [Fe/H] has been determined from Fe I lines.	
	
\section{Statistical Significance} 
\label{section5}

\citet{Fernandez-Trincado2019d} already had some indication for the existence of a distinct stellar sub-population in the inner halo of the MW, which is separated relatively cleanly in the [Al/Fe]--[Si/Fe] plane as shown in their Figure 2. This new sub-population was proven to be statistically significant and to belong to a true low-density valley separating the Si-normal population from the Si-rich population, which exceeds the background level by a factor of $\sim$4. 

\begin{figure}
		\begin{center}
			\includegraphics[width=95mm]{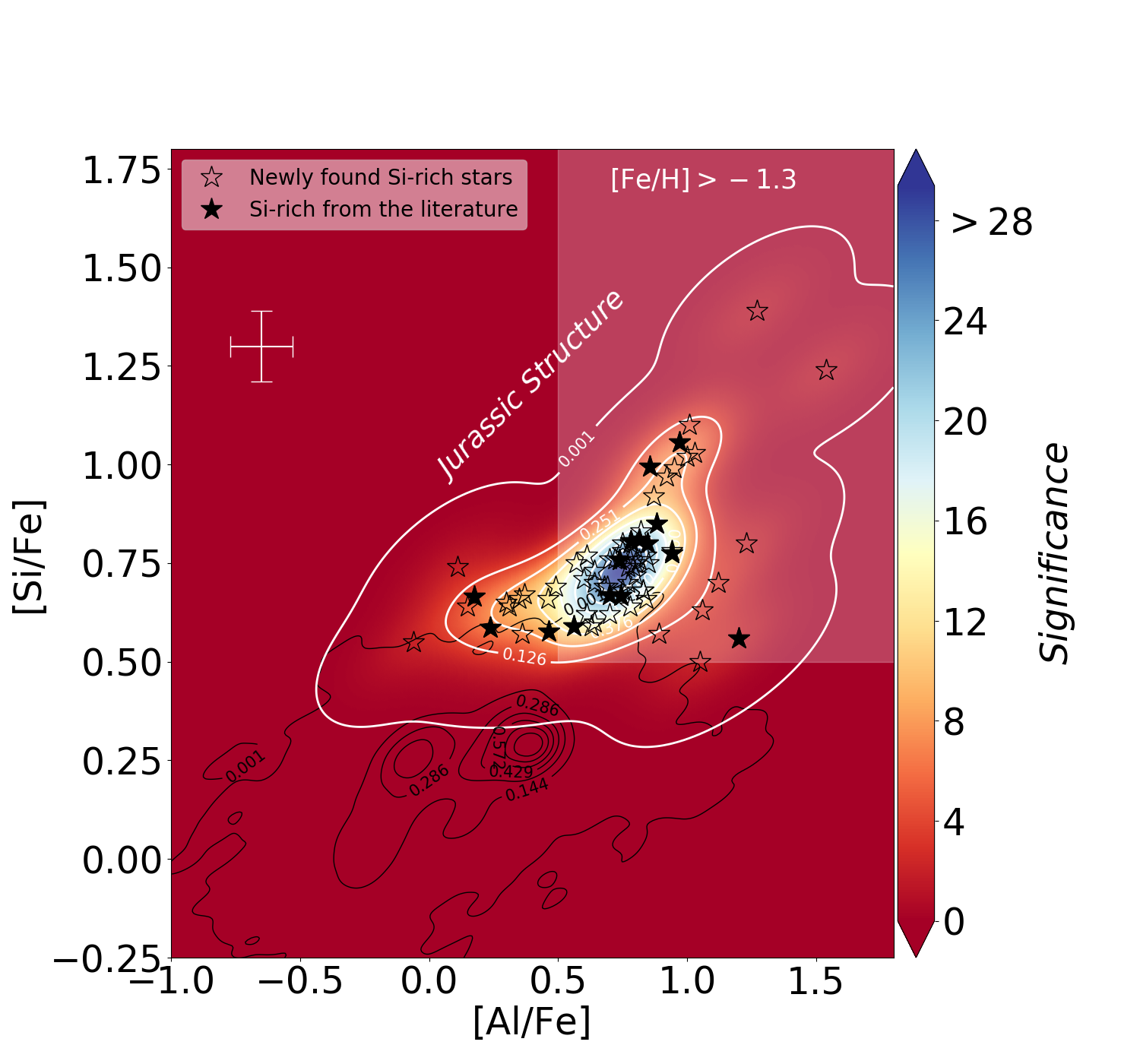}
			\caption{Distribution of [Si/Fe] vs. [Al/Fe] for the APOGEE-$2+$ stars surviving the quality cuts discussed in Section \ref{section3}.  Isophotes corresponding to the chemical domain of the relatively more metal-poor thick disk and canonical/accreted halo populations are also superimposed, while individual stars are plotted where APOGEE-$2+$ observed stars are less populous in this plane.}
			\label{Figure4}
		\end{center}
\end{figure}

With our large sample, we revisit the statistical significance of the \texttt{Jurasicc} structure over a wide range of metallicities for an unprecedented homogeneous dataset, together with other previously identified Si-rich stars from the literature and highlighted as black filled "star" symbols in Figure \ref{Figure4}. There are also one N-rich star (TYC 5619-109-1, a possible early-AGB star) analysed in \citet[][]{Fernandez-Trincado2016b} and  \citet{Pereira2017}; one N-rich bulge giant from \citet{Schiavon2017b}; one N-/Al-enhanced giant from \citet{Fernandez-Trincado2017a}; four N-rich giants from \citet{Fernandez-Trincado2019c}; and eleven Si-/Al-enhanced giants from \citet{Fernandez-Trincado2019d}. All of them (with the exception of one star from Schiavon's sample and a few N-rich stars from the Fern\'andez-Trincado's study) exhibit typical enhancement in Al (e.g., [Al/Fe]$> +$0.5), well-above the typical Galactic levels. 

Note that in \citet{Fernandez-Trincado2019d} we searched for Si-stars exclusively enriched in Al ([Al/Fe]$\gtrsim+0.5$), as aluminum appears to be one of the most effective chemical tags for GC-like abundance patterns among metal-poor stars (such a large enrichment in Al has not been observed in dwarf galaxy stellar populations; see \citealt{Shetrone2003, Hasselquist2017}). Here, we relax this restriction in order to include those mildly metal-poor giants with moderate enrichment in Al, as illustrated in Figure \ref{Figure4}. 

With this large sample, we find that $\gtrsim$80\% of the newly identified stars in the \texttt{Jurassic} structure display an aluminum enrichment ([Al/Fe] $> +$ 0.5) above the Galactic levels, while $\sim20\%$ of stars in our sample lie in a group with $-0.1\lesssim$[Al/Fe]$\lesssim +0.5$, which is likely part of the [Al/Fe]--[Si/Fe] tail of the distribution of stars  in the \texttt{Jurassic} structure, extending approximately from sub-Solar to super-Solar [Al/Fe], as seen in Figure \ref{Figure4}.

As before, we ran a Kernel Density Estimation (KDE) model over the stars  in the \texttt{Jurassic} structure, and compared them with the KDE distribution of the Si-normal giants belonging to the main body of the MW (see Figure \ref{Figure4}). The \texttt{Jurassic} structure is centred at around \{[Al/Fe],[Si/Fe]\}$=$\{$+0.75$,$+0.74$\}, while the Si-normal (MW halo and disk system) stars run roughly between \{[Al/Fe],[Si/Fe]\}$\sim$ \{$-0.26$,$+0.20$\} and \{$+0.19$,$+0.25$\}. Nevertheless, close inspection of Figure \ref{Figure4} reveals a fairly clear clump of giants (the \texttt{Jurassic} structure) which is not located on the main bulk of the KDE of the MW sample, and well-separated from the main body, exceeding the background level by a factor of $\sim$20--28 (five times more significant that determined previously). A set of white contour lines is provided as a visual aid. The  [Si,Al/Fe]-peak $\gtrsim +$0.5 is clearly visible in Figure \ref{Figure4}, which corresponds to the stars in the \texttt{Jurassic} structure of predominantly more metal-rich ([Fe/H]$\gtrsim -1.3$) stars enriched in Si and Al, compared to their extended tail, which extends from super-Solar to sub-Solar Al ([Al/Fe] $<+$0.5), and having [Fe/H]$<-$1.3 extending down to [Fe/H]$\sim-$1.8. The majority of stars in our final sample set lie in a group with super-Solar [Al/Fe] and [Si/Fe], which makes them unlikely to be field stars chemically tagged as migrants from dwarf galaxies. However, it is very likely that the parent systems were predominantly metal-rich, [Fe/H]$>-1.3$. This result confirms and reinforces the existence of a new stellar sub-population in the inner stellar halo of the MW, which is clearly well-separated from the normal halo and disk system.
	
\section{Elemental Abundance Analysis}
	\label{section6}		
		
The results of the derived elemental abundances are shown in Figures \ref{Figure5}--\ref{Figure51}. For all the chemical species we have employed the \texttt{BACCHUS} code, and manually tweaked the [X/Fe] and [Fe/H] abundance of each line until the synthetic profile matched the observed profile. The results are compared to GC stars from \citet{Meszaros2020} spanning the same metallicity range as our sample. 
	
\subsection{The iron-peak element: Fe}

The \texttt{Jurassic} structure spans a wide range in metallicity, of $\sim0.8$ dex, with two apparent peaks at [Fe/H]$=-1.2$ and $-0.9$, and an extended tail to lower metallicity ([Fe/H]$\lesssim-1.5$), which provides further evidence for several progenitors being responsible for producing the anomalous abundance ratios of [Si/Fe].  This could explain the disparate regions of the chemical space (see Figures \ref{Figure5}) and integrals of motion space (see Figure \ref{Figure6}) described below. 

\subsection{The light-elements: C and N}

Figures \ref{Figure5} show that carbon and nitrogen span similar large ranges in [C/Fe]$=\{-0.31, +0.67\}$ and [N/Fe]$=\{-0.11, +1.33\}$, with a few ($\sim$14\%) stars being slightly enhanced in carbon ([C/Fe]$\gtrsim+0.2$), along with simultaneous enrichment in the \textit{s}-process elements ([Ce, Nd/Fe]$\gtrsim{+0.3, +1.3}$), contributing to the subclass of the carbon-enhanced metal-poor \textit{s}-process-enriched (CEMP-$s$) stars at similar metallicity \citep{Carollo2014, Beers2017}, but displaying similar star-to-star scatter as that seen in Galactic GC stars at similar metallicity \citep{Meszaros2015, Masseron2019, Meszaros2020}. It is worth mentioning that a possible cause for the large dispersion observed in [C/Fe] and [N/Fe] could partially be attributed to the sensitivity of these elements to the atmospheric parameters. This could be due to the molecular equilibria that exist in the stellar atmosphere between $^{16}$OH, $^{12}$C$^{16}$O, and $^{12}$C$^{14}$N, as the strengths of the molecular features are strongly dependent on the surface temperature ($T_{\rm eff}$) and gravity ($\log$ \textit{g}), in particular for the warmer metal-poor giant stars. 

\subsection{The $\alpha$-elements: O and Mg}

Figures \ref{Figure5} show that our [O/Fe] ratios exhibit evidence of an oxygen-poor ([O/Fe]$\lesssim+0.5$) and oxygen-rich ([O/Fe]$\gtrsim+0.5$) sequence in both the intermediate and low-[Fe/H] regime, likely indicating a distinct formation history for each sequence. The stars in the \texttt{Jurassic} structure in the oxygen-poor sequence are slightly more enhanced than the canonical components of the MW, but both span similar ranges as that observed in GC stars at similar metallicity. 

From Figure \ref{Figure5}, we also note that the peculiar Si-enhanced stars in GCs \citep{Meszaros2020} occupy similar loci in [O/Fe] as those populated by our sample. The remaining $\alpha$-element abundance (Mg) appears to be very mixed, making it difficult to disentangle a clear $\alpha$-poor sequence from an $\alpha$-rich one, which could be a result of the O and Mg SNII yields being mass dependent \citep[see, e.g.,][]{Nomoto2013, Hawkins2015}.

We also find a star (2M22375002$-$1654304) in the \texttt{Jurassic} structure whose chemistry is consistent with a genuine \textit{second-generation} GCs. 2M22375002$-$1654304 is a not carbon-enhanced metal-poor star ([Fe/H]$\sim-1.27$) which has a [Mg/Fe] ratio of $\sim-1$ accompained by a modest enrichment in [N, Al, Si/Fe]$\gtrsim+0.5$, which is the typical signature of \textit{second-generation} GC stars \citep[see, e.g.,][]{Pancino2017, Masseron2019, Meszaros2020}.  It is support with the hypothesis that the \texttt{Jurassic} structure could be made up of dissipated GCs.

\subsection{The odd-Z elements: Na and Al}

We find a noticeable trend of [Al/Fe] with metallicity (see Figure \ref{Figure3}); the more metal-poor stars in our sample exhibit a low aluminum enrichment ([Al/Fe]$\lesssim+0.5$), making them equally probable to be associated with dwarf galaxy stars and/or Galactic GCs, while the more metal-rich  ([Fe/H]$\gtrsim-1.2$) stars in our sample are generally more enriched in aluminum than typically seen in dwarf galaxy stars \citep[see, e.g.,][]{Hasselquist2017, Hayes2018,  Hasselquist2019, Nidever2020}.  These stars are more likely linked to a GC origin, and span a similar [Al/Fe] spread with an apparent weak N-Al correlation (see Figure \ref{Figure51}--\textit{top}) as that observed in metal-rich/-poor GC stars \citep{Meszaros2020}. 

From Figure \ref{Figure51}--\textit{bottom}, we also note two trends of [Al/Fe] with [Mg/Fe]; the stars in the \texttt{Jurassic} structure with low-[Al/Fe] ([Al/Fe]$\lesssim+0.5$) stars follow an apparent Al-Mg anti-correlation as that seen in Galactic GCs \citep{Meszaros2020}. However, an unexpected turnover in the Al-Mg diagram is also clearly seen in Figure \ref{Figure3}, where the aluminum-enriched stars in our sample  ([Al/Fe]$\gtrsim+0.5$) display an apparent Al-Mg correlation, with the apparent existence of an Al-Si correlation from the data in Table \ref{Table1}. This could be interpreted as a signature of $^{28}$Si leakage from the MgAl chain \citep{Masseron2019, Szabolcs2020} in their different progenitors, likely happening toward higher metallicities ([Fe/H]$\gtrsim-1.2$). It is important to note, however, that the $^{28}$Si leakage has been only observed thus far in metal-poor ([Fe/H]$\lesssim-1.2$) GCs \citep[see, e.g.,][]{Masseron2019, Szabolcs2020}, while the stars  in the \texttt{Jurassic} structure are particularly more metal rich. 

Whether some of the metal-rich stars  in the \texttt{Jurassic} structure were part of the GC stars that have fully migrated into the inner halo of the MW could in part explain their lack in metal-rich GCs, but it is still unclear. However, our finding may indicate that the temperature conditions of the stars in the progenitors were too high to efficiently produce the $^{28}$Si leakage from Mg-Al chain \citep{Prantzos2017, Masseron2019} toward higher metallicities. 

Figure \ref{Figure5} shows the trends of sodium as a function of metallicity for our sample. The abundance of Na exhibits a small dispersion ($<$0.22 dex) and remains rather constant at [Fe/H]$\gtrsim-1.2$, except for one metal-poor star in our sample that exhibits a high enrichment in [Na/Fe]$\gtrsim+0.8$. However, it is important to note that [Na/Fe] abundances have been found to be typically affected by large uncertainties ($\gtrsim0.2$ dex), as its two lines (1.6373$\mu$m and 1.6388$\mu$m) in our APOGEE-$2+$ spectra are weak (line intensity is comparable to the variance) and possibly blended at the typical $T_{\rm eff}$ and metallicity of our sample, which would lead to unreliable abundance results. For this reason, the [Na/Fe] listed in Table \ref{Table2} are listed as upper limits.

\subsection{The \textit{s}-process elements: Ce and Nd}
	
\begin{figure*}
\begin{center}
\includegraphics[width=210mm]{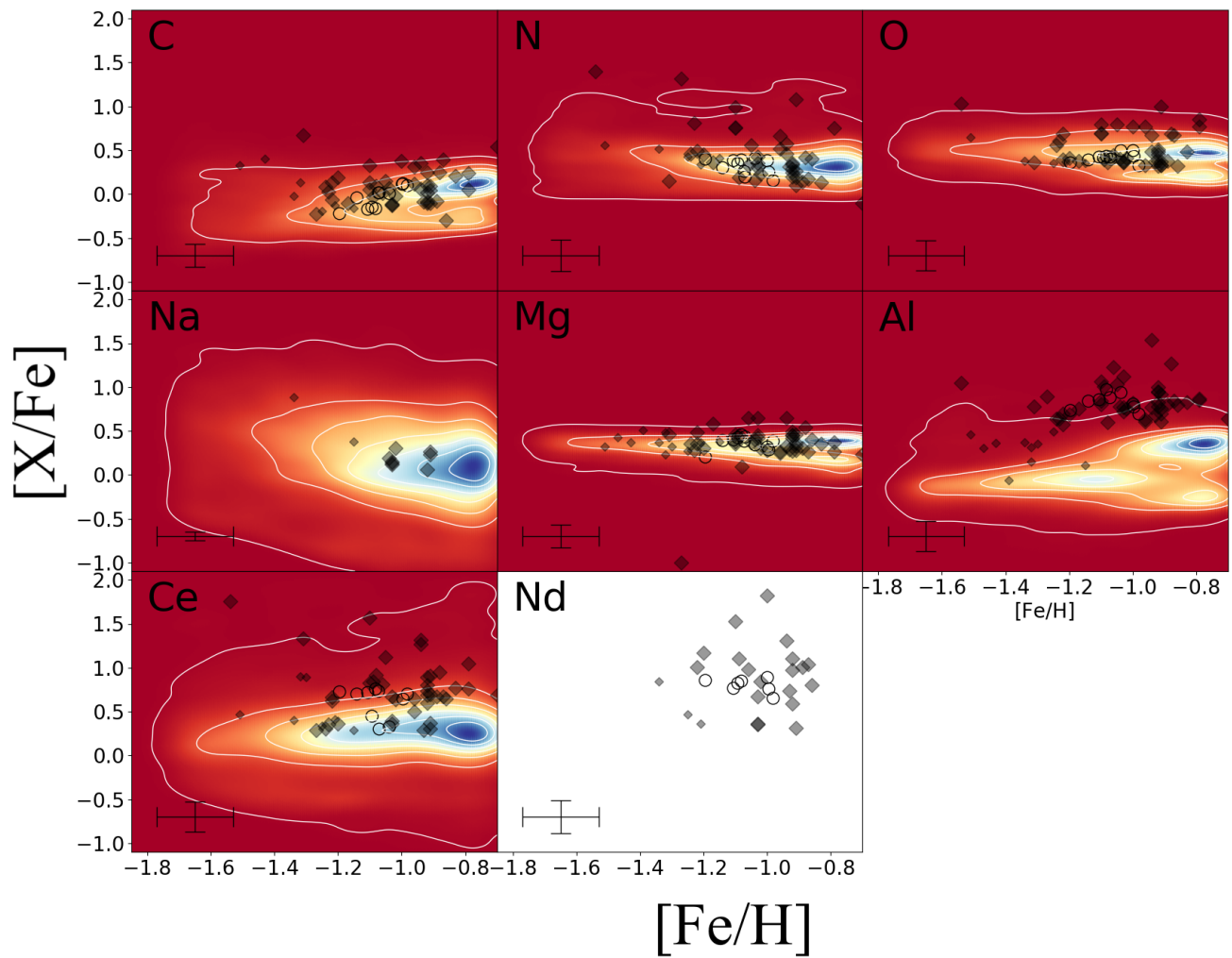}
\caption{Kernel Density Estimation (KDE) of [X/Fe] with metallicity for the APOGEE-2$+$ stars surviving the quality cuts discussed in Section \ref{section3}. Stars in the \texttt{Jurassic} structure are shown as black diamonds. The small diamond symbols mark the Si-rich stars with [Al/Fe]$<+0.5$, while the large diamond symbols refer to Si-rich stars with [Al/Fe]$\gtrsim+0.5$. Previously discovered Si-rich stars from \citet{Fernandez-Trincado2019d} are highlighted with black open circles.}
\label{Figure5}
\end{center}
\end{figure*}

\begin{figure}
	\begin{center}
		\includegraphics[width=90mm]{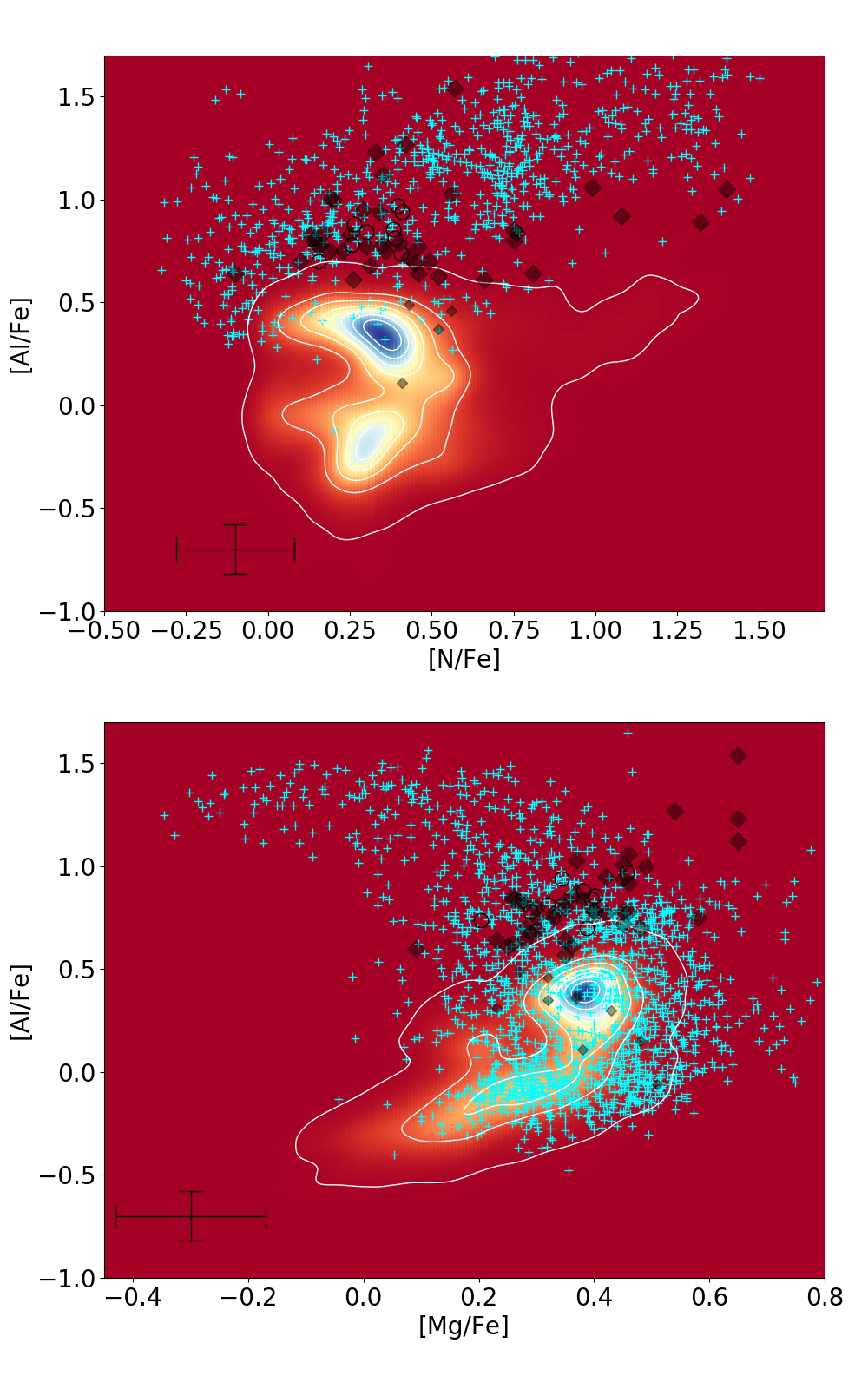}
		\caption{Same as Figure \ref{Figure5}, but for abundance distributions of [Al/Fe] with [N/Fe] (\textit{top}) and [Mg/Fe] (\textit{bottom}). Over-plotted (cyan crosses) are GCs from \citet{Meszaros2020} at similar metallicity as the Si-rich stars.}
		\label{Figure51}
	\end{center}
\end{figure}

The \texttt{Jurassic} structure exhibits a clear enhancement of the heavy second \textit{s}-process-peak elements, such as Ce and Nd, as shown in Figures \ref{Figure5}, with [Ce/Fe]$=+0.29$ to $+1.76$ and [Nd/Fe]$=+0.31$ to $+1.82$. Depending on the chemical composition of the other chemical species, the stars in the \texttt{Jurassic} structure could owe itheir heavy-element abundance patterns to different possible channels, likely by contamination from mass transfer \citep[see, e.g.,][]{Preston2001, Lucatello2003, Sivarani2004, Barbuy2005, Thompson2008, Fernandez-Trincado2019b}, or by pollution of gas already strongly enriched in \textit{s}-process elements. 

The modest carbon enhancement observed in a handful ~($\sim14\%$ of the stars in our sample) of stars in the \texttt{Jurassic} structure may raise the possibility that other enrichment processes, rather than the mass-transfer hypothesis, could be responsible for the over-abundance in \textit{s}-process elements. These abundance signatures are often observed among chemically peculiar giant stars, CEMP-$s$ stars, and low-/intermediate-mass asymptotic giant branch (AGB) stars at this metallicity \citep[see, e.g.,][]{Carollo2014, Fernandez-Trincado2016b, Ventura2016a, Beers2017, Pereira2017, Pereira2019}, or by pollution of nearby massive (10--300 M$_{\odot}$) stars \citep[see, e.g.,][]{Masseron2020}.  Additionally, there is good agreement in the [Ce, Nd/Fe] abundance ratios derived in this work with the \textit{s}-process elements in some GC stars \citep{Masseron2019, Szabolcs2020}.
	
\section{Orbits}
\label{section7}

We used the \texttt{GravPot16}\footnote{\url{https://gravpot.utinam.cnrs.fr}} code to study the Galactic orbits of the stars  in the \texttt{Jurassic} structure. For this purpose, we combine accurate proper motions from \textit{Gaia} DR2 \citep{Brown2018, Lindegren2018}, radial velocities from APOGEE-$2+$ \citep{Nidever2015, Majewski2017, Ahumada2020}, and distances from \texttt{StarHorse} \citep{Anders2019, Queiroz2018, Queiroz2020a, Queiroz2020b} as input data in our model. Only stars with a re-normalized unit weight error \citep{Lindegren2018}, \texttt{RUWE}$<1.4$ (stars with reliable proper motions) were considered in our orbital analysis. 
	
The Galactic potential assumed in these calculations is the non-axisymmetric MW-like potential, which considers perturbations due to a rotating boxy/peanut bar.  This potential fits the structural and dynamical parameters of the Galaxy, based on recent knowledge of our MW. For each star, we computed an ensemble of orbits by assuming three different values for the angular velocity of the bar, $\Omega_{bar} =$ 33, 43, and 53 km s$^{-1}$ kpc$^{-1}$, with a bar mass of 1.1$\times10^{10}$ M$_{\odot}$, and a present-day angle orientation of 20$^{\circ}$, in the same manner as in \citet{Fernandez-Trincado2020a}. To model the uncertainty distributions, we sampled one million orbits using a simple Monte Carlo approach assuming Gaussian distributions for the input parameters (heliocentric distances, radial velocities, and proper motions), with $1\sigma$ equal to the errors of the input parameters as listed in Table \ref{Table1}. The main orbital elements ($|Z_{max}|$, $r_{peri}$, $r_{apo}$, eccentricity, and the z-component of the angular momentum in the inertial frame) are listed in Table \ref{Table5}. The data presented in this table correspond to a backward time integration of 3 Gyr, with error bars computed as $\Delta=$ (84$^{\rm th}$ percentile $- 16^{\rm th}$ percentile)/2, while the number inside parenthesis indicate the sensitivity in the orbital elements due to the variations of the angular velocity of the bar. This table also list the minimum and maximum variation of the z-component of the angular momentum in the inertial frame, $L_{z}$, since this quantity is not conserved in a model like \texttt{GravPot16}, and we are interested in the variation of $L_{z}$ along the full integration time, allowing us to identify the orbital configuration of each star: prograde, retrograde, and P-R\footnote{A P-R orbit is defined as to the one that flip its sense from prograde to retrograde, or vice-versa, along its orbit.} Orbits are calculated with respect to the rotation of the bar.
	
For reference, the Galactic convention adopted by this work is: $X-$axis is oriented toward $l=$ 0$^{\circ}$ and $b=$ 0$^{\circ}$, and the $Y-$axis is oriented toward $l$ = 90$^{\circ}$ and $b=$0$^{\circ}$, and the disk rotates toward $l=$ 90$^{\circ}$; the velocities are also oriented in these directions. In this convention, the Sun's orbital velocity vector is [U$_{\odot}$,V$_{\odot}$,W$_{\odot}$] = [$11.1$, $12.24$, 7.25] km s$^{-1}$ \citep{Brunthaler2011}. The model has been rescaled to the Sun's Galactocentric distance, 8 kpc, and a local rotation velocity of $V_{LSR} = 244.5$ km s$^{-1}$ \citep{Fernandez-Trincado2020a}.
    
We find that most of the stars in the \texttt{Jurassic} structure span a large range in heliocentric distances, $ 3 < d < 12$ kpc, are more likely found at intermediate to high latitudes $|b| > 10^{\circ}$, with a few exceptions to the inner regions of the MW, and currently are located far from the peri-/apocentre of their orbits.
    
The orbital elements reveal that most ($\sim 70\%$) of the stars in the \texttt{Jurassic} structure have highly eccentric orbits ($e \gtrsim 0.7$), covering a wide range of vertical excursions from the Galactic plane ($\lesssim46$ kpc).  These appear to behave as halo-like orbits, some of which have \textit{mid-} and \textit{off-plane} orbits passing through the inner Galaxy. A small fraction of the stars in the \texttt{Jurassic} structure have orbits concentrated at small heights above the plane, with apocentric distances, $r_{apo}$, that vary between 4.5 kpc to 13 kpc, and pericenter distances between $\sim$3 kpc to 6 kpc from the Galactic Centre, with almost circular prograde orbits, suggesting a possible dynamical association with the thick disk. 

\begin{figure*}
	\begin{center}
		\includegraphics[width=180mm]{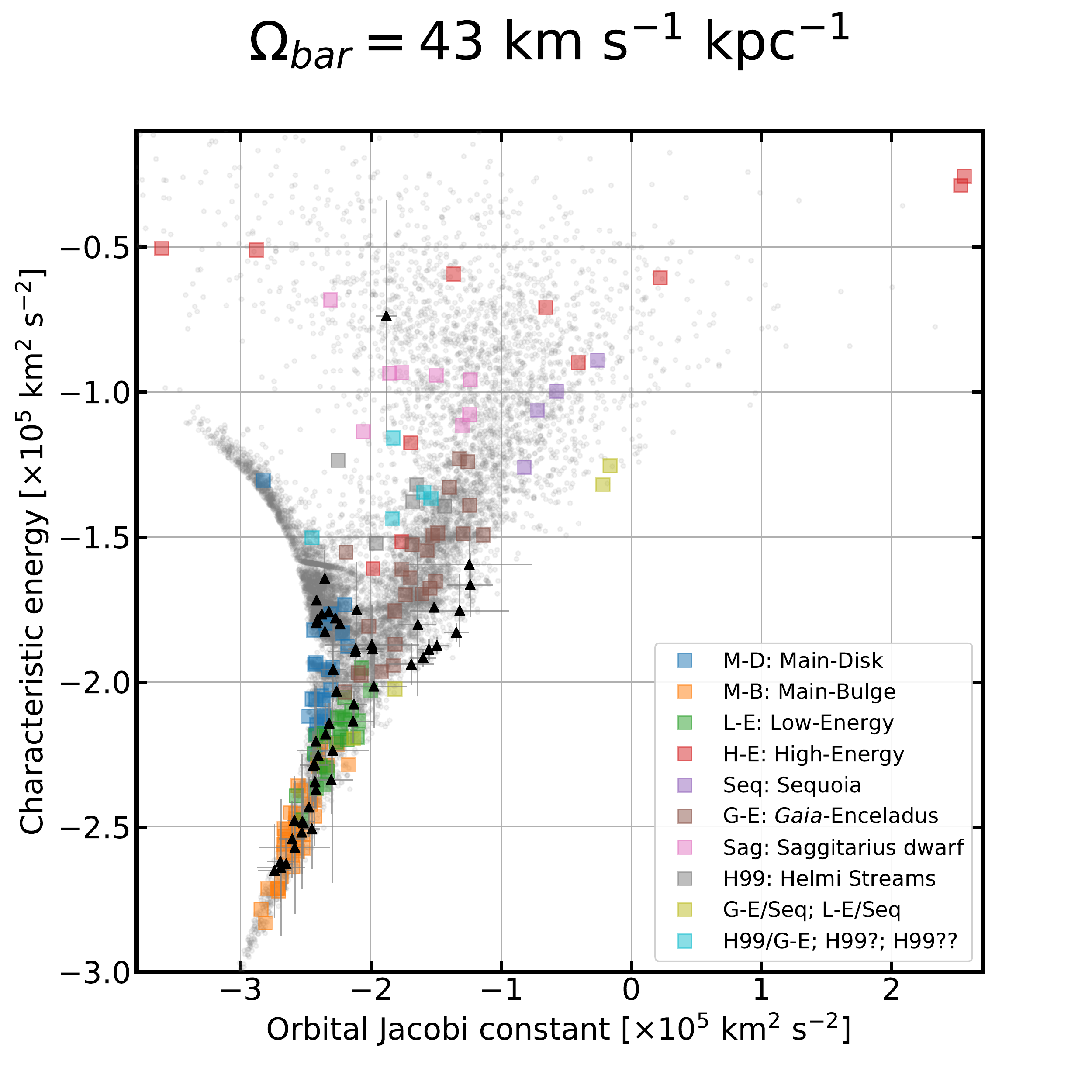}
		\caption{Characteristic orbital energy (($E_{max} + E_{min}$)/2) versus the orbital Jacobi constant ($E_{J}$) in the non-inertial reference frame where the bar is at rest. Square symbols refer to Galactic GCs, colour-coded according to their associations with different progenitors from \citep{Massari2019}. The black triangles with error bars refer to the stars  in the \texttt{Jurassic} structure analysed in this study. The grey dots show the APOGEE-2$+$ stars surviving the quality cuts discussed in Section \ref{section3}.}
		\label{Figure6}
	\end{center}
\end{figure*}

For comparison with our sample, we calculated the orbital solutions for Galactic GCs from \citet{Baumgardt2019}, and adopted the progenitor classification as in \citet{Massari2019}. Figure \ref{Figure6} shows the characteristic orbital energy (($E_{max} + E_{min}$)/2) versus the orbital Jacobi constant ($E_{J}$) distribution of Galactic GCs and the stars in the \texttt{Jurassic} structure. This diagram clearly shows that the stars  in the \texttt{Jurassic} structure populate a wide range of energies, similar to that of Galactic GCs with different origins, suggesting that the stars in the \texttt{Jurassic} structure are peculiar objects that appear not to have emerged from a single system.

\begin{figure}
	\begin{center}
		\includegraphics[width=90mm]{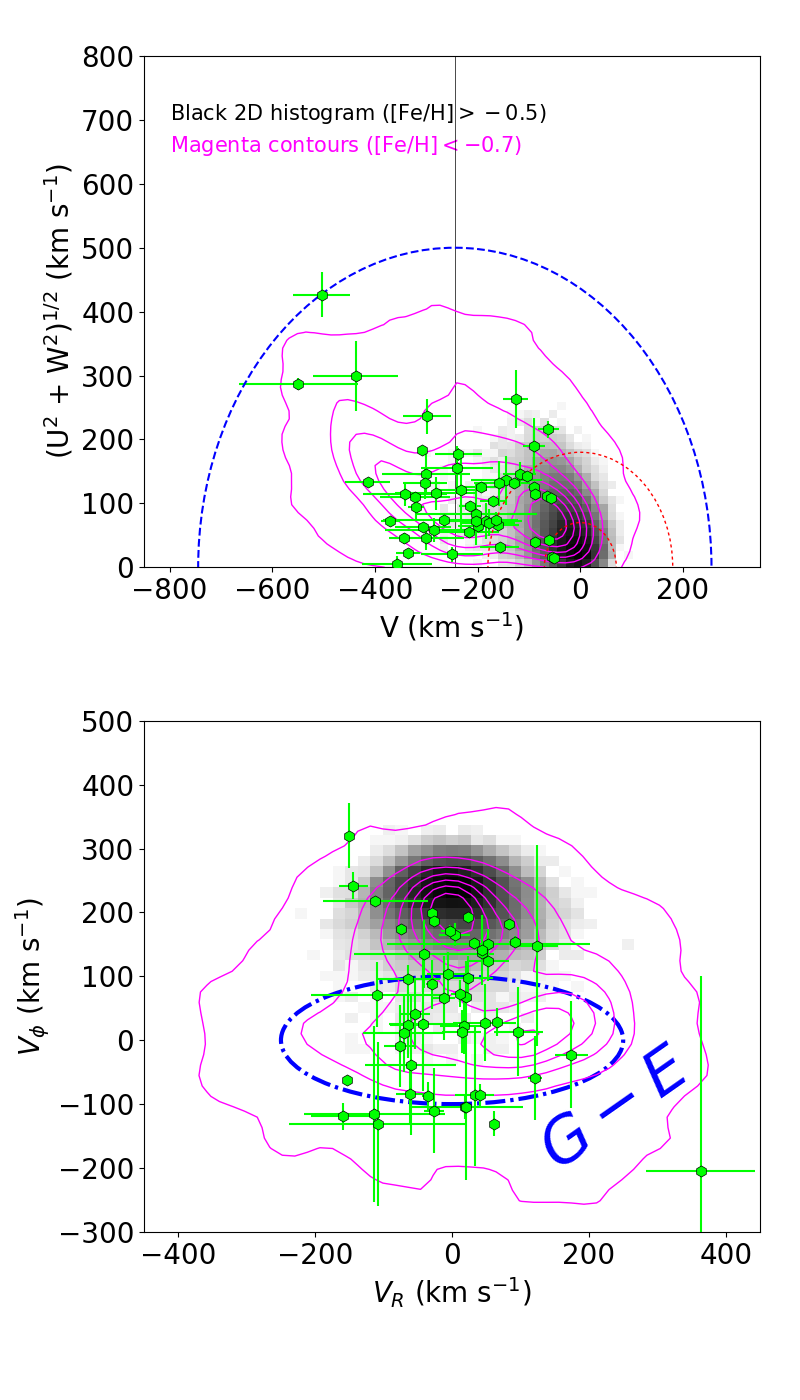}
		\caption{Toomre diagram (\textit{top}) for the stars in the \texttt{Jurassic} structure (lime symbols) compared to a control sample of $\sim$200,000 APOGEE-2$+$ stars with [Fe/H]$>-0.5$ and $d_{\odot} < 5$ kpc, likely associated with the thin and thick disk (2D histrogram), and a sample of $\sim$ 11,000 metal-poor stars, [Fe/H]$<-0.5$, likely associated with the halo and thick disk population (magenta contours). All velocities are relative to the LSR. The red dashed rings show roughly the boundaries of the thin and thick disk at a constant velocity of $\sim$70 km s$^{-1}$ and 180 km s$^{-1}$, respectively \citep{Venn2004}. The blue dashed ring represents a constant galactic-rest frame velocity of $\sim$500 km s$^{-1}$ shifted relative to the other velocities. The distribution of the velocity components $V_{\phi}$ vs. $V_{R}$ (\textit{bottom}) is also shown for our sample. The blue dashed ellipse represents the approximate region for stars associated with the \textit{Gaia}-Enceladus in $V_{\phi}$ vs. $V_{R}$ space based on \citet{Belokurov2018}.}
		\label{Figure7}
	\end{center}
\end{figure}

Figure \ref{Figure7} show the distribution of stars in the \texttt{Jurassic} structure in the Toomre diagram and $V_{\phi}$ vs. $V_{R}$ space. In these plane, we can see that the vast majority of the stars in the \texttt{Jurassic} structure stand out in their distribution of velocity components and are clearly distinct from the disk, with a few exceptions associated with stars having low orbital eccentricities as listed in Table \ref{Table5}, which have been likely trapped in corotation resonance with the bar. The $V_{\phi}$ vs. $V_{R}$ plane, reveals that some of the stars in the \texttt{Jurassic} structure are associated with the \textit{Gaia}-Enceladus as also revealed by their orbital elements as seen in Figure \ref{Figure6}. It is likely that the \texttt{Jurassic} structure is made up of the cumulative effect of several tidally disrupted GCs, some of which belonged to the \textit{Gaia}-Enceladus accretion event.

\subsection{Results of the Orbital Analysis}
	
From the energy plane (see Figure \ref{Figure6}), we can clearly distinguish three possible groups among the stars in the \texttt{Jurassic} structure. The stars in this structure with $E_{J} \lesssim -2\times10^5$ km$^{2}$ s$^{-2}$ populate the region dominated by \textit{in-situ} clusters belonging to the subgroup of GCs in the Main-Disk (M-D), Main-Bulge (M-B), and Low-Energy (L-E) group \citep[see, e.g.,][]{Massari2019}, while one star  in the \texttt{Jurassic} structure has an $E_{J} \lesssim -2\times10^5$ km$^{2}$ s$^{-2}$, occupying the region populated by GCs belonging the high-energy (H-E) group and those associated with the \texttt{Sagittarius} (Sgr) dwarf spheroidal galaxy, with some contamination by clusters associated with the GSE progenitor(s) \citep[see, e.g.,][]{Belokurov2018, Massari2019}. 
		 
Strikingly, we also find a clear substructure of 10 stars in the \texttt{Jurassic} structure with $E_{J} \gtrsim -2\times10^5$ km$^{2}$ s$^{-2}$, which appear contained within the region dominated by the subgroup of GCs associated with GSE \citep[see, e.g.,][]{Massari2019}. This particular subgroup of stars in the \texttt{Jurassic} structure ~correspond mostly with the stars with a low aluminum enrichment ([Al/Fe]$\lesssim+0.5$), with retrograde orbits, eccentricities ranging from $e \sim0.47$ to 0.9, vertical excursions from the Galactic plane from 1.56 kpc to 7.5 kpc, $r_{peri} = $ 0.7 -- 3.13 kpc, and $r_{apo} = $ 7.8--15.36 kpc. Thus, given the low [Al/Fe] of some (6 out of 10) stars in this subgroup, they are likely associated with the accreted GSE progenitor(s) \citep[e.g.,][]{Belokurov2018, Helmi2018, Haywood2018, Myeong2019}, or possibly are linked to GC debris from the GSE merger event. For the remaining stars in the \texttt{Jurassic} structure with high-[Al/Fe] it is unlikely that they came from dwarf galaxy progenitors, as present-day stars in dwarf galaxy satellites do not exhibit [Al/Fe] abundance ratios which exceed $\gtrsim+0.5$, thus the more probable origin of those stars may be associated with \textit{in-situ} and Low-Energy Galactic GCs, with a few exceptions, as the case of the stars in the \texttt{Jurassic} structure with orbital energies in the domain of the Sgr clusters (see Figure \ref{Figure6}). 
		 	
\section{Concluding Remarks}
\label{section8}
	
Using the internal APOGEE-$2+$ data set, we confirm and reinforce the existence of a distinct stellar population in the inner halo of the MW--the \texttt{Jurassic} structure. Here, we report on the identification of 55 new Si-rich, mildly metal-poor stars, 45 of which exhibit high [Al/Fe]$\gtrsim+0.5$, making it unlikely that dwarf galaxy satellites could have contributed the majority of these stars. We find some dynamical evidencethat a few of them, in particular the stars  in the \texttt{Jurassic} structure with [Al/Fe]$\lesssim+0.5$, could be associated with the accreted GSE dwarf galaxy and/or GCs of the GSE progenitor(s). 
	
We also find that the large majority of the new stars in the \texttt{Jurassic} structure exhibit chemical and dynamical properties similar to that of \textit{in-situ} GCs, supporting the picture that these peculiar stars have likely been dynamically ejected into the inner halo by dissolved GCs. This discovery could aid in explaining the assembly of the mildly metal-poor ([Fe/H]$\sim-1$) component of the inner Galactic halo and its complex formation process.
	
	\begin{acknowledgements}  
		The author is grateful for the enlightening feedback from the anonymous referee. J.G.F-T is supported by FONDECYT No. 3180210. T.C.B. acknowledges partial support for this work from grant PHY 14-30152: Physics Frontier Center / JINA Center for the Evolution of the Elements (JINA-CEE), awarded by the US National Science Foundation. D.M. is supported by the BASAL Center for Astrophysics and Associated Technologies (CATA) through grant AFB 170002, and by project FONDECYT Regular No. 1170121.\\

        This work has made use of data from the European Space Agency (ESA) mission Gaia (\url{http://www.cosmos.esa.int/gaia}), processed by the Gaia Data Processing and Analysis Consortium (DPAC, \url{http://www.cosmos.esa.int/web/gaia/dpac/consortium}). Funding for the DPAC has been provided by national institutions, in particular the institutions participating in the Gaia Multilateral Agreement.\\

        Funding for the Sloan Digital Sky Survey IV has been provided by the Alfred P. Sloan Foundation, the U.S. Department of Energy Office of Science, and the Participating Institutions. SDSS- IV acknowledges support and resources from the Center for High-Performance Computing at the University of Utah. The SDSS web site is www.sdss.org. SDSS-IV is managed by the Astrophysical Research Consortium for the Participating Institutions of the SDSS Collaboration including the Brazilian Participation Group, the Carnegie Institution for Science, Carnegie Mellon University, the Chilean Participation Group, the French Participation Group, Harvard-Smithsonian Center for Astrophysics, Instituto de Astrof\`{i}sica de Canarias, The Johns Hopkins University, Kavli Institute for the Physics and Mathematics of the Universe (IPMU) / University of Tokyo, Lawrence Berkeley National Laboratory, Leibniz Institut f\"{u}r Astrophysik Potsdam (AIP), Max-Planck-Institut f\"{u}r Astronomie (MPIA Heidelberg), Max-Planck-Institut f\"{u}r Astrophysik (MPA Garching), Max-Planck-Institut f\"{u}r Extraterrestrische Physik (MPE), National Astronomical Observatory of China, New Mexico State University, New York University, University of Notre Dame, Observat\'{o}rio Nacional / MCTI, The Ohio State University, Pennsylvania State University, Shanghai Astronomical Observatory, United Kingdom Participation Group, Universidad Nacional Aut\'{o}noma de M\'{e}xico, University of Arizona, University of Colorado Boulder, University of Oxford, University of Portsmouth, University of Utah, University of Virginia, University of Washington, University of Wisconsin, Vanderbilt University, and Yale University.\\
	\end{acknowledgements}


\clearpage	
\newpage	
	
\begin{appendix}
	
\section{Tables}

We provide our results in five tables. The first table, a sample of which can be found in Table \ref{Table1}, contains the basic parameters of stars in the \texttt{Jurassic} structure, i.e., 2MASS magnitudes, atmospheric parameters and radial velocity from the APOGEE$2+$ survey, astrometric information from \texttt{Gaia} DR2. Table \ref{Table2}, contains the final abundances for every element (C, N, O, Na, Mg, Al, Si, Fe, Ce, and Nd) and star on a line-by-line basis as analyzed in this work with the \texttt{BACCHUS} code (see Section \ref{abundances} for more details), while the Tables \ref{Table3} and \ref{Table4}, contains the uncertainties in the abundance caused by uncertainties in the stellar parameters for a small fraction of stars in our sample. 

Table \ref{Table5}, contains the main orbital elements of stars in the \texttt{Jurassic} structure predicted with the \texttt{GravPot16} code (see Section \ref{section7} for more details).    
	
		\begin{sidewaystable*}
			\begin{tiny}
				\setlength{\tabcolsep}{0.9mm}  
				\caption{Basic parameters of stars in the \texttt{Jurassic} structure.}
				\centering
				\begin{tabular}{|c|ccccccccccccrrrccc|}
					\hline
					APOGEE ID        & $J-K_{s}$  & E(B-V)  & $A_{K_s}^{WISE}$ & $A_{K_s}^{TARG}$&  [M/H] &  $T^{ASPCAP}_{\rm eff}$ & $\log$ \textit{g}$^{ASPCAP}$ & $T^{pho}_{\rm eff}$ & log \textit{g}$^{iso}$ & N$_{\rm visits}$  & S/N &  $RV_{SCATTER}$  & \textit{RV} &  $\mu_{\alpha}\cos(\delta)$ & $\mu_{\delta}$ & d$_{\odot}$ & Telescope & \texttt{RUWE} \\
					& 2MASS  &   & & &   &  uncalibrated &  uncalibrated & & &   &  &   &&  & & &  &  \\
					&   &  &   & &   &  [K] & [cgs] & [K] & [cgs] &   & [pixel$^{-1}$] &  [km s$^{-1}$] & [km s$^{-1}$] & [mas yr$^{-1}$] & [mas yr$^{-1}$] & [kpc] &  & \\
					\hline
					\hline
					2M09133506$+$2248579 &  0.435 &  0.042 &    0.011  &  0.011  &  $-$1.38 & 5429 &  2.39 &  5549 &  3.49 &  6  &  126   &  0.63  &    $+$190.6   &  $-$3.02 $\pm$ 0.05 &   $-$10.16 $\pm$ 0.03 &  5.97 $\pm$  1.51 &  APO 2.5m  &  1.21 \\
					2M12242950$+$4408525 &  0.355 &  0.009 &    0.007  &  0.002  &  $-$1.47 & 5524 &  3.46 &  5859 &  3.66 &  2  &  133   &  0.03  &   $-$7.3        & $-$12.32 $\pm$ 0.03 &  $-$169.85 $\pm$ 0.03 &  0.45 $\pm$  0.01 &  APO 2.5m  &  1.05 \\
					2M12443130$-$0900220 &  0.367 &  0.034 &    0.006  &  0.014  &  $-$1.39 & 5469 &  3.40 &  5873 &  3.69 &  3  &  126   &  0.33  &    $+$58.2     & $-$58.11 $\pm$ 0.10 &   $+$12.73 $\pm$ 0.08 &  0.45 $\pm$  0.01 &  APO 2.5m  &  1.09 \\
					2M13472354$+$2210562 &  0.391 &  0.018 &    0.065  &  0.005  &  $-$1.31 & 5375 &  2.36 &  5696 &  3.63 &  3  &  164   &  0.47  &    $+$30.8     & $-$20.59 $\pm$ 0.05 &   $-$12.11 $\pm$ 0.05 &  2.17 $\pm$  0.21 &  APO 2.5m  &  1.10 \\
					2M13535604$+$4437076 &  0.506 &  0.008 &    0.033  &  0.038  &  $-$0.91 & 5146 &  2.81 &  5384 &  3.52 &  3  &  346   &  0.01  & $-$128.9      & $-$44.35 $\pm$ 0.03 &   $-$31.06 $\pm$ 0.04 &  1.44 $\pm$  0.08 &  APO 2.5m  &  0.99 \\
					2M14533964$+$4506180 &  0.526 &  0.018 &    0.039  &  0.005  &  $-$1.29 & 5021 &  2.78 &  5149 &  2.60 &  3  &   87   &  0.03  & $-$162.2       & $-$21.85 $\pm$ 0.02 &   $-$13.50 $\pm$ 0.02 &  2.64 $\pm$  0.18 &  APO 2.5m  &  1.03 \\
					2M15170852$+$4033475 &  0.359 &  0.017 &    0.026  &  0.005  &  $-$1.43 & 5674 &  2.31 &  5852 &  3.68 &  1  &  113   &  ...   & $-$171.0         & $-$20.42 $\pm$ 0.03 &   $-$18.25 $\pm$ 0.04 &  2.34 $\pm$  0.21 &  APO 2.5m  &  1.04 \\
					2M15275895$+$4226412 &  0.857 &  0.023 &    0.090  &  0.007  &  $-$1.22 & 4137 &  1.23 &  4169 &  0.78 &  5  &  614   &  0.08  & $-$285.1      &  $-$7.57 $\pm$ 0.03 &   $-$11.79 $\pm$ 0.04 &  5.16 $\pm$  0.83 &  APO 2.5m  &  0.99 \\
					2M16013102$+$0618450 &  0.620 &  0.048 &    0.080  &  0.014  &  $-$1.21 & 4805 &  2.38 &  4865 &  2.06 &  7  &  127   &  0.29  & $-$104.4      &  $-$3.74 $\pm$ 0.04 &    $-$5.92 $\pm$ 0.03 &  6.44 $\pm$  1.81 &  APO 2.5m  &  1.02 \\
					2M16441013$-$1850478 &  1.022 &  0.632 &    0.120  &  0.120  &  $-$0.88 & 4590 &  2.23 &  4150 &  0.97 &  5  &  108   &  0.09  &     $+$17.5    &  $-$2.11 $\pm$ 0.09 &    $-$2.92 $\pm$ 0.05 &  9.94 $\pm$  2.14 &  LCO 2.5m  &  1.01 \\
					2M16595910$+$1127496 &  0.661 &  0.059 & $-$0.012  &  0.017  &  $-$0.70 & 5076 &  2.75 &  4750 &  2.25 &  24 &   95   &  0.41  &   $+$34.5   &  $-$2.67 $\pm$ 0.07 &    $-$4.84 $\pm$ 0.08 &  7.48 $\pm$  3.07 &  APO 2.5m  &  1.01 \\
					2M17255079$-$2029099 &  0.987 &  0.572 &    0.211  &  0.211  &  $-$0.90 & 4402 &  1.79 &  4559 &  1.70 &  2  &  165   &  0.26  &     $+$13.1    &  $-$5.71 $\pm$ 0.05 &    $-$7.00 $\pm$ 0.04 &  7.53 $\pm$  1.29 &  LCO 2.5m  &  0.88 \\
					2M17265466$-$1331522 &  1.096 &  0.548 &    0.177  &  0.177  &  $-$0.88 & 4383 &  2.00 &  4154 &  0.98 &  3  &  129   &  0.03  &  $-$12.2       &  $-$2.91 $\pm$ 0.08 &    $-$5.01 $\pm$ 0.06 & 10.94 $\pm$  3.52 &  LCO 2.5m  &  1.01 \\
					2M17513049$+$5801309 &  0.575 &  0.042 &    0.067  &  0.012  &  $-$1.38 & 4885 &  2.50 &  5009 &  2.21 &  1  &  201   &  ...   & $-$101.3         &  $+$3.86 $\pm$ 0.05 &    $-$6.77 $\pm$ 0.07 &  2.11 $\pm$  0.17 &  APO 2.5m  &  0.90 \\
					2M17572447$-$3056414 &  0.873 &  ...   &    ...    &  0.249  &  $-$1.22 & 4763 &  1.94 &  5085 &  2.53 &  1  &   73   &  ...   &     $+$96.5                &  $+$0.76 $\pm$ 0.08 &    $-$4.38 $\pm$ 0.07 &  8.02 $\pm$  1.99 &  LCO 2.5m  &  0.99 \\
					2M18013098$-$3307263 &  1.088 &  0.745 &    0.179  &  0.179  &  $-$0.93 & 4140 &  1.25 &  4183 &  1.03 &  1  &  136   &  ...   & $-$154.2         &  $-$1.07 $\pm$ 0.08 &    $-$7.19 $\pm$ 0.07 &  7.93 $\pm$  1.43 &  LCO 2.5m  &  0.82 \\
					2M18043255$-$4819138 &  0.932 &  0.151 &    0.123  &  0.123  &  $-$1.17 & 4174 &  1.24 &  4375 &  1.15 &  2  &  106   &  0.05  &  $-$53.0       &  $-$1.04 $\pm$ 0.05 &    $-$8.31 $\pm$ 0.05 &  7.55 $\pm$  2.00 &  LCO 2.5m  &  1.14 \\
					2M18151248$-$4403407 &  0.829 &  0.119 &    0.100  &  0.100  &  $-$1.05 & 4319 &  1.70 &  4581 &  1.60 &  4  &  161   &  0.14  &     $+$21.4    &  $-$2.92 $\pm$ 0.08 &    $-$6.80 $\pm$ 0.07 &  7.50 $\pm$  1.79 &  LCO 2.5m  &  0.99 \\
					2M18453994$-$3010465 &  1.043 &  0.153 &    0.147  &  0.144  &  $-$1.24 & 3835 &  0.31 &  4165 &  0.77 &  1  &  254   &  ...   &    $+$173.9      &  $+$2.25 $\pm$ 0.07 &    $-$4.32 $\pm$ 0.06 &  6.18 $\pm$  1.43 &  APO 2.5m  &  0.94 \\
					2M19105369$+$2717150 &  0.670 &  0.247 &    0.023  &  0.022  &  $-$0.85 & 5230 &  3.20 &  4746 &  2.05 &  7  &   73   &  0.18  &   $-$5.1        &  $-$2.85 $\pm$ 0.03 &   $-$13.04 $\pm$ 0.03 &  4.58 $\pm$  0.81 &  APO 2.5m  &  1.03 \\
					2M19193412$-$2931210 &  0.710 &  0.154 &    0.080  &  0.046  &  $-$1.15 & 4909 &  2.32 &  4720 &  1.77 &  6  &   87   &  2.34  &  $-$16.0        &  $-$5.21 $\pm$ 0.04 &    $-$3.83 $\pm$ 0.04 &  6.77 $\pm$  1.97 &  APO 2.5m  &  1.03 \\
					2M19214936$-$1232462 &  1.071 &  0.182 &    0.149  &  0.149  &  $-$1.06 & 3949 &  0.77 &  4122 &  0.78 &  3  &  265   &  0.23  & $-$146.6      &  $-$1.98 $\pm$ 0.07 &    $-$0.67 $\pm$ 0.04 &  7.61 $\pm$  2.01 &  APO 2.5m  &  0.89 \\
					2M22045404$-$1148287 &  0.725 &  0.080 &    0.044  &  0.024  &  $-$1.26 & 4435 &  1.69 &  4579 &  1.43 &  3  &  550   &  0.02  & $-$172.7      &  $-$3.88 $\pm$ 0.07 &    $-$5.96 $\pm$ 0.07 &  1.53 $\pm$  0.14 &  LCO 2.5m  &  1.27 \\
					2M16241820$-$2145485 &  1.169 &  0.428 &    0.235  &  0.129  &  $-$1.20 & 3866 &  0.50 &  ...  &  ...  &  1  &  172   &  ...   &     $+$66.4             &  $-$4.40 $\pm$ 0.09 &    $-$4.49 $\pm$ 0.05 &  6.94 $\pm$  1.34 &  LCO 2.5m  &  0.94 \\
					2M17295157$-$3737045 &  1.785 &  ...   &    0.684  &  0.653  &  $-$1.20 & 4427 &  1.51 &  ...  &  ...  &  1  &   67   &  ...   & $-$178.6                     &  $-$7.61 $\pm$ 0.32 &    $-$3.53 $\pm$ 0.25 & 16.23 $\pm$  1.54 &  LCO 2.5m  &  1.20 \\
					2M02175837$-$7313144 &  0.965 &  0.042 &    0.071  &  0.012  &  $-$0.92 & 4019 &  1.18 &  ...  &  ...  &  5  &  376   &  0.12  &     $+$47.3          &   $+$2.37 $\pm$ 0.05 &   $+$0.04 $\pm$ 0.04 &  5.01 $\pm$  0.75 &  LCO 2.5m  &  1.09 \\
					2M04463289$-$7336289 &  0.764 &  0.095 &    0.082  &  0.028  &  $-$1.03 & 4504 &  2.03 &  4490 &  1.50 &  1  &  169   &  ...   &    $+$48.2       &   $+$3.22 $\pm$ 0.04 &   $+$3.29 $\pm$ 0.05 &  3.51 $\pm$  0.46 &  LCO 2.5m  &  1.06 \\
					2M13315603$+$4354469 &  0.842 &  0.020 &    0.054  &  0.006  &  $-$1.14 & 4433 &  1.85 &  4203 &  0.92 &  1  &  207   &  ...   & $-$124.5        &  $-$9.42 $\pm$ 0.03 &   $-$11.06 $\pm$ 0.03 &  5.06 $\pm$  1.04 &  APO 2.5m  &  1.03 \\
					2M13590229$-$0831592 &  0.670 &  0.034 &    0.046  &  0.010  &  $-$0.91 & 4923 &  2.69 &  4691 &  1.95 &  6  &   95   &  0.33  &      $+$7.6     &  $-$3.04 $\pm$ 0.09 &    $-$6.12 $\pm$ 0.09 &  7.16 $\pm$  4.03 &  APO 2.5m  &  1.19 \\
					2M15160402$+$4728327 &  0.601 &  0.031 &    0.045  &  0.009  &  $-$0.94 & 5221 &  3.13 &  4904 &  2.38 &  2  &  101   &  0.01  &  $-$35.4      & $-$12.96 $\pm$ 0.02 &    $-$4.75 $\pm$ 0.03 &  1.57 $\pm$  0.07 &  APO 2.5m  &  1.07 \\
					2M16005847$-$3708334 &  0.934 &  0.468 &    0.195  &  0.195  &  $-$0.93 & 4562 &  2.24 &  4651 &  1.87 &  3  &  123   &  0.23  &  $-$41.7       &  $-$5.22 $\pm$ 0.06 &    $-$6.27 $\pm$ 0.04 &  9.00 $\pm$  2.39 &  LCO 2.5m  &  0.92 \\
					2M17073023$-$2717147 &  1.050 &  0.440 &    0.198  &  0.198  &  $-$1.14 & 4132 &  1.13 &  4335 &  1.15 &  1  &   78   &  ...   & $-$257.5          & $-$11.97 $\pm$ 0.06 &   $-$13.31 $\pm$ 0.03 &  6.84 $\pm$  1.38 &  LCO 2.5m  &  0.89 \\
					2M17171752$-$2148372 &  0.934 &  0.617 &    0.138  &  0.138  &  $-$1.05 & 4399 &  2.09 &  4429 &  1.32 &  1  &  118   &  ...   &  $-$88.6          &  $-$3.53 $\pm$ 0.06 &   $-$10.57 $\pm$ 0.04 &  6.89 $\pm$  1.63 &  APO 2.5m  &  0.96 \\
					2M18220651$-$3835466 &  1.017 &  0.162 &    0.147  &  0.147  &  $-$1.05 & 3946 &  0.54 &  4245 &  1.00 &  2  &  183   &  0.02  &  $-$42.8       &  $-$0.97 $\pm$ 0.05 &    $-$5.80 $\pm$ 0.04 &  6.64 $\pm$  1.54 &  LCO 2.5m  &  1.15 \\
					2M18242843$-$3942097 &  0.671 &  0.107 &    0.048  &  0.048  &  $-$1.23 & 4693 &  2.23 &  4848 &  2.02 &  4  &   91   &  0.11  &  $-$54.2        &  $-$1.37 $\pm$ 0.06 &    $-$9.40 $\pm$ 0.06 &  7.02 $\pm$  1.89 &  LCO 2.5m  &  0.96 \\
					2M18280709$-$3900094 &  0.724 &  0.087 &    0.038  &  0.038  &  $-$0.93 & 4415 &  2.01 &  4646 &  1.86 &  2  &   97   &  0.40  &     $+$73.6     &  $-$1.66 $\pm$ 0.05 &    $-$6.13 $\pm$ 0.04 &  7.08 $\pm$  1.79 &  LCO 2.5m  &  1.24 \\
					2M22184840$-$0431434 &  0.667 &  0.072 &    0.063  &  0.021  &  $-$0.89 & 4858 &  2.72 &  4748 &  2.05 &  2  &  192   &  0.04  &  $-$27.4       &   $+$1.41 $\pm$ 0.08 &    $-$4.84 $\pm$ 0.13 &  2.43 $\pm$  0.32 &  APO 2.5m  &  1.32 \\
					2M22375002$-$1654304 &  0.599 &  0.044 &    0.121  &  0.013  &  $-$1.21 & 4784 &  1.84 &  4933 &  2.20 &  1  &  181   &  ...   &   $-$4.3           &   $+$3.53 $\pm$ 0.07 &    $-$9.94 $\pm$ 0.07 &  4.20 $\pm$  1.55 &  APO 2.5m  &  0.92 \\
					2M17401617$-$2806344 &  2.055 &  ...   &    ...    &  0.697  &  $-$1.02 & 4195 &  1.35 &  ...  &  ...  &  2  &  100   &  0.12  &  $-$80.5                      &  $-$3.86 $\pm$ 0.78 &    $-$2.39 $\pm$ 0.65 & ...  &  LCO 2.5m  &  1.29 \\
					2M14513934$-$0602148 &  0.541 &  0.066 &    0.140  &  0.020  &  $-$1.40 & 4933 &  2.36 &  5171 &  2.59 &  27 &   96   &  0.94  &     $+$90.9   &  $-$4.88 $\pm$ 0.07 &    $-$6.77 $\pm$ 0.07 &  6.82 $\pm$  3.06 &  APO 2.5m  &  0.96 \\
					2M22480199$+$1411329 &  0.614 &  0.044 &    0.060  &  0.013  &  $-$1.18 & 4842 &  2.51 &  4879 &  2.08 &  3  &  251   &  0.34  & $-$210.9     &  $+$20.44 $\pm$ 0.06 &   $-$22.99 $\pm$ 0.05 &  1.91 $\pm$  0.18 &  APO 2.5m  &  1.04 \\
					2M16544476$-$3939140 &  1.343 &  ...   &    0.369  &  0.534  &  $-$1.05 & 4798 &  2.50 &  4793 &  1.99 &  4  &   74   &  0.15  &     $+$67.7        &  $-$3.17 $\pm$ 0.39   &    $-$5.82 $\pm$ 0.26 &  9.85 $\pm$  2.43 &  LCO 2.5m  &  1.04 \\
					2M20440538$-$0713572 &  0.574 &  0.057 &    0.046  &  0.017  &  $-$1.09 & 5106 &  2.94 &  5035 &  2.50 &  4  &  136   &  0.03  & $-$151.8     & $-$35.42 $\pm$ 1.03  &   $-$45.51 $\pm$ 0.84 &  3.07 $\pm$  1.38 &  APO 2.5m  & 21.69 \\
					2M13435919$-$1807213 &  0.736 &  0.096 &    0.053  &  0.029  &  $-$1.01 & 4947 &  2.78 &  4574 &  1.65 &  1  &   77   &  ...   &    $+$117.8      &  $-$5.66 $\pm$ 0.04  &   $-$14.66 $\pm$ 0.04 &  4.29 $\pm$  0.73 &  APO 2.5m  &  1.17 \\
					2M17183459$+$4302520 &  0.596 &  0.021 &    0.001  &  ...    &  $-$1.03 & 4993 &  2.93 &  4878 &  2.24 &  15 &   90   &  0.27  & $-$135.8         &  $-$3.91 $\pm$ 0.04  &    $-$4.26 $\pm$ 0.04 &  8.39 $\pm$  3.04 &  APO 2.5m  &  0.97 \\
					2M18091354$-$2810087 &  1.329 &  ...   &    0.168  &  0.168  &  $-$1.06 & 3877 &  0.73 &  ...  &  ...  &  2  &  233   &  0.09  &     $+$87.4              &  $-$8.48 $\pm$ 0.08  &    $-$4.79 $\pm$ 0.06 &  8.08 $\pm$  1.23 &  LCO 2.5m  &  0.92 \\
					2M13303961$+$2719096 &  0.569 &  0.009 &    0.018  &  0.002  &  $-$0.94 & 5086 &  2.96 &  4984 &  2.54 &  3  &  150   &  1.49  &     $+$45.2  & $-$13.06 $\pm$ 0.03  &    $-$1.74 $\pm$ 0.01 &  2.39 $\pm$  0.20 &  APO 2.5m  &  0.96 \\
					2M15243300$+$2819313 &  0.708 &  0.028 &    0.045  &  0.008  &  $-$0.93 & 4508 &  2.18 &  4576 &  1.73 &  2  &  222   &  0.16  & $-$165.5     &  $-$3.85 $\pm$ 0.03  &    $-$7.15 $\pm$ 0.04 &  3.52 $\pm$  0.71 &  APO 2.5m  &  0.93 \\
					2M16130340$-$3144580 &  0.685 &  0.227 &    0.086  &  0.086  &  $-$1.04 & 4986 &  2.57 &  4969 &  2.43 &  2  &   83   &  0.09  &     $+$32.3    &  $-$2.24 $\pm$ 0.08  &    $-$6.16 $\pm$ 0.06 &  6.52 $\pm$  1.89 &  LCO 2.5m  &  1.05 \\
					2M16483594$-$0150117 &  0.763 &  0.185 &    0.047  &  0.047  &  $-$0.92 & 4639 &  2.34 &  4566 &  1.72 &  1  &  110   &  ...   &      $+$8.8       &  $-$7.25 $\pm$ 0.04  &    $-$6.74 $\pm$ 0.03 &  3.99 $\pm$  0.99 &  LCO 2.5m  &  1.04 \\
					2M17161376$-$2910175 &  1.069 &  0.677 &    0.169  &  0.212  &  $-$0.96 & 4150 &  1.62 &  4339 &  1.23 &  2  &  108   &  0.03  &    $+$109.6  & $-$10.13 $\pm$ 0.48 &    $-$4.39 $\pm$ 0.33 &  9.27 $\pm$  1.54 &  APO 2.5m  &  5.46 \\
					2M17171046$-$3007398 &  1.393 &  ...   &    0.388  &  0.388  &  $-$0.96 & 4108 &  1.38 &  4125 &  0.85 &  3  &   98   &  0.24  & $-$112.7           &  $-$3.83 $\pm$ 0.13  &    $-$3.84 $\pm$ 0.08 &  8.49 $\pm$  1.77 &  LCO 2.5m  &  0.90 \\
					2M17421220$-$3443594 &  1.388 &  ...   &    0.387  &  0.387  &  $-$0.97 & 4443 &  1.94 &  4131 &  0.86 &  5  &  105   &  0.23  &     $+$27.3       & $-$11.18 $\pm$ 0.17 &    $-$5.23 $\pm$ 0.14 &  9.15 $\pm$  1.71 &  LCO 2.5m  &  0.86 \\
					2M18024132$-$2940238 &  0.971 &  ...   &    ...    &  0.181  &  $-$0.98 & 4645 &  2.28 &  4490 &  1.50 &  6  &  109   &  0.11  & $-$141.8             &  $-$1.99 $\pm$ 0.20  &    $-$2.45 $\pm$ 0.16 &  9.45 $\pm$  2.08 &  LCO 2.5m  &  0.98 \\
					2M16543450$-$0429397 &  0.687 &  0.265 &    0.078  &  0.078  &  $-$0.87 & 5117 &  2.73 &  4928 &  2.43 &  4  &  110   &  0.19  &  $-$40.4      &  $+$3.34 $\pm$ 0.05  &    $-$7.07 $\pm$ 0.03 &  5.79 $\pm$  1.34 &  LCO 2.5m  &  1.02 \\
					\hline
				\end{tabular}  \label{Table1}
			\end{tiny}
		\end{sidewaystable*}

		\begin{sidewaystable*}
			\begin{tiny}
				\setlength{\tabcolsep}{1.0mm}  
				\caption{Elemental abundances of stars in the \texttt{Jurassic} structure.}
				\centering
				\begin{tabular}{|c|cccccccccc|cccccccccc|}
					\hline
        APOGEE ID            & [C/Fe]$^{*}$  &  [N/Fe]$^{*}$ & [O/Fe]$^{*}$  & [Na/Fe]$^{*}$ & [Mg/Fe]$^{*}$ & [Al/Fe]$^{*}$ & [Si/Fe]$^{*}$ & [Fe/H]$^{*}$  & [Ce/Fe]$^{*}$ & [Nd/Fe] $^{*}$& [C/Fe]$^{\dagger}$  & [N/Fe]$^{\dagger}$  & [O/Fe]$^{\dagger}$  & [Na/Fe]$^{\dagger}$ & [Mg/Fe]$^{\dagger}$ & [Al/Fe]$^{\dagger}$ & [Si/Fe]$^{\dagger}$ & [Fe/H]$^{\dagger}$  & [Ce/Fe]$^{\dagger}$ & [Nd/Fe]$^{\dagger}$ \\
					\hline
					\hline
2M09133506$+$2248579 &    ...  &    ...  & ...     & ...     & $+$0.48 & $+$0.15 & $+$0.64 & $-$1.32 & ...     &  ...    &    ...  &    ...  & ...     &  ...    & $+$0.27 & $+$0.01 & $+$0.59 & $-$1.27 &    ...  & ...     \\
2M12242950$+$4408525 &    ...  &    ...  & ...     & ...     & $+$0.43 & $+$0.30 & $+$0.65 & $-$1.47 & ...     &  ...    &    ...  &    ...  & ...     &  ...    & $+$0.46 & $+$0.46 & $+$0.65 & $-$1.34 &    ...  & ...     \\
2M12443130$-$0900220 & $+$0.40 &    ...  & ...     & ...     & $+$0.37 & $+$0.36 & $+$0.57 & $-$1.43 & ...     &  ...    & $+$0.13 &    ...  & ...     &  ...    & $+$0.43 & $+$0.50 & $+$0.59 & $-$1.26 &    ...  & ...     \\
2M13472354$+$2210562 & $+$0.13 &    ...  & ...     & ...     & $+$0.23 & $+$0.31 & $+$0.64 & $-$1.32 & $+$0.90 &  ...    & $+$0.27 &    ...  & ...     &  ...    & $+$0.15 & $+$0.51 & $+$0.61 & $-$1.14 & $+$1.43 & ...     \\
2M13535604$+$4437076 & $+$0.25 & $+$0.42 & $+$0.69 & ...     & $+$0.54 & $+$1.27 & $+$1.39 & $-$0.88 & $+$0.95 &  ...    & $+$0.35 & $+$0.92 & $+$0.54 &  ...    & $+$0.29 & $+$1.09 & $+$1.26 & $-$0.78 & $+$1.34 & ...     \\
2M14533964$+$4506180 &    ...  &    ...  & ...     & ...     & $+$0.32 & $+$0.35 & $+$0.66 & $-$1.30 & $+$0.89 &  ...    &    ...  &    ...  & ...     &  ...    & $+$0.41 & $+$0.65 & $+$0.64 & $-$1.22 & $+$0.51 & ...     \\
2M15170852$+$4033475 &    ...  &    ...  & ...     & ...     & $+$0.51 & $-$0.06 & $+$0.55 & $-$1.39 &    ...  &  ...    &    ...  &    ...  & ...     &  ...    & $+$0.27 & $-$0.01 & $+$0.52 & $-$1.29 &    ...  & ...     \\
2M15275895$+$4226412 & $-$0.13 & $+$0.34 & $+$0.37 & ...     & $+$0.37 &    ...  & $+$0.73 & $-$1.21 & $+$0.41 & $+$0.36 & $-$0.23 & $+$0.49 & $+$0.49 &  ...    & $+$0.48 &    ...  & $+$0.67 & $-$1.25 & $+$0.29 & $+$0.27 \\
2M16013102$+$0618450 & $+$0.03 & $+$0.81 & $+$0.61 & ...     & $+$0.28 & $+$0.64 & $+$0.73 & $-$1.23 & $+$0.35 &  ...    & $-$0.10 & $+$0.61 & $+$0.69 &  ...    & $+$0.39 & $+$0.69 & $+$0.70 & $-$1.17 & $+$0.25 & ...     \\
2M16441013$-$1850478 & $+$0.25 & $+$0.34 & $+$0.51 & ...     & $+$0.45 & $+$0.94 & $+$0.78 & $-$0.92 & $+$0.89 & $+$1.10 & $+$0.10 & $+$0.14 & $+$0.15 &  ...    & $+$0.57 & $+$0.75 & $+$0.87 & $-$1.21 & $+$0.35 & $+$0.68 \\
2M16595910$+$1127496 & $+$0.54 & $-$0.10 &    ...  & ...     & $+$0.23 & $+$0.64 & $+$0.60 & $-$0.70 & $+$0.69 &  ...    & $+$0.38 & $-$0.28 &    ...  &  ...    & $+$0.29 & $+$0.30 & $+$0.70 & $-$0.85 & $+$0.41 & ...     \\
2M17255079$-$2029099 & $-$0.30 & $+$0.40 & $+$0.32 & ...     & $+$0.40 & $+$0.79 & $+$0.74 & $-$0.86 & $+$0.65 & $+$0.80 & $-$0.36 & $+$0.46 & $+$0.54 &  ...    & $+$0.53 & $+$0.91 & $+$0.65 & $-$0.79 & $+$0.62 & $+$0.75 \\
2M17265466$-$1331522 & $+$0.06 & $+$0.57 & $+$0.48 & ...     & $+$0.65 & $+$1.54 & $+$1.24 & $-$0.94 & $+$1.32 & $+$1.31 & $-$0.10 & $+$0.58 & $+$0.29 &  ...    & $+$0.93 & $+$1.57 & $+$1.18 & $-$1.12 & $+$0.68 & $+$0.91 \\
2M17513049$+$5801309 & $+$0.33 & $+$0.56 & $+$0.65 & ...     & $+$0.32 & $+$0.46 & $+$0.66 & $-$1.51 & $+$0.47 &  ...    & $+$0.32 & $+$0.77 & $+$0.78 &  ...    & $+$0.41 & $+$0.52 & $+$0.62 & $-$1.46 & $+$0.28 & ...     \\
2M17572447$-$3056414 & $+$0.13 &    ...  & $+$0.69 & ...     & $+$0.35 & $+$0.57 & $+$0.75 & $-$1.22 & $+$0.62 &  ...    & $+$0.01 & $+$0.77 &    ...  &  ...    & $+$0.29 & $+$0.63 & $+$0.82 & $-$1.05 & $+$0.70 & ...     \\
2M18013098$-$3307263 & $-$0.10 & $+$0.26 & $+$0.33 & $+$0.26 & $+$0.36 & $+$0.61 & $+$0.77 & $-$0.91 & $+$0.30 & $+$0.31 & $-$0.19 & $+$0.32 & $+$0.40 & $+$0.23 & $+$0.52 & $+$0.72 & $+$0.83 & $-$0.87 & $+$0.23 & $+$0.27 \\
2M18043255$-$4819138 & $-$0.10 & $+$0.36 & $+$0.38 & ...     & $+$0.58 & $+$0.75 & $+$0.80 & $-$1.17 &    ...  &  ...    & $-$0.16 & $+$0.53 & $+$0.69 &  ...    & $+$0.69 & $+$0.92 & $+$0.67 & $-$1.09 &    ...  & ...     \\
2M18151248$-$4403407 & $+$0.07 & $+$0.16 & $+$0.40 & $+$0.19 & $+$0.40 & $+$0.78 & $+$0.76 & $-$1.04 &    ...  &  ...    & $-$0.01 & $+$0.41 & $+$0.76 & $+$0.03 & $+$0.56 & $+$0.96 & $+$0.65 & $-$0.95 &    ...  & ...     \\
2M18453994$-$3010465 & $-$0.02 & $+$0.52 & $+$0.38 & $+$0.88 & $+$0.37 & $+$0.37 & $+$0.67 & $-$1.34 & $+$0.40 & $+$0.84 & $+$0.04 & $+$0.61 & $+$0.88 & $+$0.89 & $+$0.35 & $+$0.67 & $+$0.52 & $-$1.20 & $+$0.72 & $+$1.08 \\
2M19105369$+$2717150 & $+$0.24 & $+$0.75 & $+$0.85 & ...     & $+$0.26 & $+$0.85 & $+$0.75 & $-$0.79 & $+$1.05 &  ...    & $+$0.05 & $+$0.68 & $+$0.69 &  ...    & $+$0.52 & $+$0.99 & $+$0.87 & $-$1.03 & $+$0.46 & ...     \\
2M19193412$-$2931210 & $+$0.05 & $+$0.76 & $+$0.69 & ...     & $+$0.38 & $+$0.84 & $+$0.77 & $-$1.10 & $+$0.83 &  ...    & $+$0.01 & $+$0.66 & $+$0.58 &  ...    & $+$0.48 & $+$0.87 & $+$0.78 & $-$1.20 & $+$0.34 & ...     \\
2M19214936$-$1232462 & $-$0.12 & $+$0.30 & $+$0.36 & $+$0.12 & $+$0.42 & $+$0.77 & $+$0.74 & $-$1.03 & $+$0.39 & $+$0.36 & $-$0.17 & $+$0.43 & $+$0.68 & $+$0.11 & $+$0.53 & $+$0.98 & $+$0.68 & $-$0.95 & $+$0.54 & $+$0.51 \\
2M22045404$-$1148287 & $+$0.67 & $+$0.15 & $+$0.36 & ...     & $+$0.46 & $+$0.78 & $+$0.80 & $-$1.31 & $+$1.33 &  ...    & $+$0.60 & $+$0.54 & $+$0.48 &  ...    & $+$0.59 & $+$0.88 & $+$0.72 & $-$1.28 & $+$1.19 & ...     \\
2M16241820$-$2145485 & $-$0.19 & $+$0.43 & $+$0.35 & ...     & $+$0.27 & $+$0.49 & $+$0.69 & $-$1.25 & $+$0.34 & $+$0.47 &    ...  &    ...  &    ...  &  ...    &    ...  &    ...  &    ...  & ...     &    ...  & ...     \\
2M17295157$-$3737045 & $-$0.06 & $+$0.49 & $+$0.39 & ...     & $+$0.48 & $+$0.70 & $+$0.76 & $-$1.22 & $+$0.66 & $+$1.01 &    ...  &    ...  &    ...  &  ...    &    ...  &    ...  &    ...  & ...     &    ...  & ...     \\
2M02175837$-$7313144 & $-$0.09 & $+$0.20 & $+$0.38 & $+$0.06 & $+$0.49 & $+$1.00 & $+$1.02 & $-$0.92 & $+$0.69 &  ...    &    ...  &    ...  &    ...  &  ...    &    ...  &    ...  &    ...  & ...     &    ...  & ...     \\
2M04463289$-$7336289 & $-$0.02 & $+$0.23 & $+$0.41 & $+$0.30 & $+$0.30 & $+$0.75 & $+$0.78 & $-$1.02 & $+$0.64 & $+$0.84 & $-$0.14 & $+$0.36 & $+$0.45 & $+$0.32 & $+$0.47 & $+$0.85 & $+$0.76 & $-$1.05 & $+$0.36 & $+$0.49 \\
2M13315603$+$4354469 & $-$0.03 & $+$0.19 & $+$0.39 & ...     & $+$0.45 & $+$1.01 & $+$1.10 & $-$1.09 & $+$0.84 & $+$1.10 & $-$0.17 & $+$0.21 & $+$0.22 &  ...    & $+$0.59 & $+$1.02 & $+$0.94 & $-$1.27 & $+$0.34 & $+$0.53 \\
2M13590229$-$0831592 & $+$0.39 & $+$0.17 & $+$0.67 & ...     & $+$0.26 & $+$0.84 & $+$0.66 & $-$0.87 & $+$0.67 & $+$1.04 & $+$0.24 & $+$0.16 & $+$0.56 &  ...    & $+$0.45 & $+$0.19 & $+$0.69 & $-$0.98 & $+$0.39 & $+$0.80 \\
2M15160402$+$4728327 &    ...  & $+$1.08 & $+$1.00 & ...     & $+$0.46 & $+$0.92 & $+$0.97 & $-$0.91 & $+$0.91 &  ...    &    ...  & $+$1.03 & $+$0.70 &  ...    & $+$0.66 & $+$1.05 & $+$0.84 & $-$1.07 & $+$0.42 & ...     \\
2M16005847$-$3708334 & $+$0.02 & $+$0.29 & $+$0.45 & ...     & $+$0.42 & $+$0.95 & $+$0.99 & $-$0.92 & $+$0.80 & $+$0.97 & $-$0.11 & $+$0.35 & $+$0.58 &  ...    & $+$0.64 & $+$1.11 & $+$0.98 & $-$0.88 & $+$0.64 & $+$0.81 \\
2M17073023$-$2717147 & $-$0.16 & $+$0.41 & $+$0.35 & $+$0.38 & $+$0.38 & $+$0.11 & $+$0.74 & $-$1.15 & $+$0.29 &  ...    & $-$0.19 & $+$0.56 & $+$0.67 & $+$0.36 & $+$0.50 & $+$0.66 & $+$0.64 & $-$1.06 & $+$0.35 & ...     \\
2M17171752$-$2148372 & $+$0.09 &    ...  & $+$0.36 & ...     & $+$0.09 & $+$0.60 & $+$0.71 & $-$1.08 & $+$0.92 &  ...    & $-$0.14 &    ...  & $+$0.46 &  ...    & $+$0.33 & $+$0.80 & $+$0.69 & $-$1.10 & $+$0.51 & ...     \\
2M18220651$-$3835466 & $-$0.13 & $+$0.43 & $+$0.36 & $+$0.13 & $+$0.45 & $+$0.72 & $+$0.76 & $-$1.03 & $+$0.36 & $+$0.35 & $-$0.08 & $+$0.68 & $+$0.84 & $+$0.04 & $+$0.50 & $+$0.94 & $+$0.67 & $-$0.87 & $+$0.67 & $+$0.73 \\
2M18242843$-$3942097 & $+$0.08 & $+$0.46 & $+$0.56 & ...     & $+$0.35 & $+$0.64 & $+$0.70 & $-$1.24 & $+$0.30 &  ...    & $-$0.02 & $+$0.67 & $+$0.73 &  ...    & $+$0.49 & $+$0.72 & $+$0.65 & $-$1.17 & $+$0.19 & ...     \\
2M18280709$-$3900094 & $-$0.08 & $+$0.14 & $+$0.38 & ...     & $+$0.30 & $+$0.80 & $+$0.72 & $-$0.89 & $+$0.67 & $+$1.01 & $-$0.18 & $+$0.42 & $+$0.65 &  ...    & $+$0.51 & $+$0.99 & $+$0.64 & $-$0.81 & $+$0.64 & $+$0.95 \\
2M22184840$-$0431434 & $+$0.21 & $+$0.13 & $+$0.48 & ...     & $+$0.35 & $+$0.83 & $+$0.68 & $-$0.83 & $+$0.77 &  ...    & $+$0.10 & $+$0.15 & $+$0.45 &  ...    & $+$0.58 & $+$0.98 & $+$0.70 & $-$0.89 & $+$0.35 & ...     \\
2M22375002$-$1654304 & $-$0.23 & $+$1.32 &    ...  & ...     & $-$1.00 & $+$0.89 & $+$0.57 & $-$1.27 & $+$0.29 &  ...    & $+$0.27 & $+$0.82 &    ...  &  ...    & $-$1.01 & $+$0.91 & $+$0.56 & $-$1.20 & $+$0.38 & ...     \\
2M17401617$-$2806344 & $+$0.01 & $+$0.33 & $+$0.44 & ...     & $+$0.65 & $+$1.23 & $+$0.80 & $-$1.06 & $+$0.81 & $+$0.98 &    ...  &    ...  &    ...  &  ...    &    ...  &    ...  &    ...  & ...     &    ...  & ...     \\
2M14513934$-$0602148 &    ...  & $+$1.40 & $+$1.03 & ...     &    ...  & $+$1.05 & $+$0.50 & $-$1.54 & $+$1.76 &  ...    &    ...  & $+$1.55 & $+$0.98 &  ...    &    ...  & $+$1.12 & $+$0.55 & $-$1.41 & $+$1.65 & ...     \\
2M22480199$+$1411329 & $+$0.19 & $+$0.45 & $+$0.41 & ...     & $+$0.30 & $+$0.69 & $+$0.69 & $-$1.20 & $+$0.36 & $+$1.17 & $+$0.19 & $+$0.50 & $+$0.49 &  ...    & $+$0.47 & $+$0.75 & $+$0.67 & $-$1.21 & $+$0.22 & $+$0.92 \\
2M16544476$-$3939140 & $+$0.13 & $+$0.99 & $+$0.70 & ...     & $+$0.46 & $+$1.06 & $+$0.63 & $-$1.10 & $+$1.57 & $+$1.53 & $+$0.03 & $+$1.07 & $+$0.79 &  ...    & $+$0.69 & $+$1.21 & $+$0.60 & $-$1.10 & $+$1.30 & $+$1.31 \\
2M20440538$-$0713572 & $+$0.33 & $+$0.75 & $+$0.79 & ...     & $+$0.35 & $+$0.80 & $+$0.75 & $-$1.10 &    ...  &  ...    & $+$0.23 & $+$0.83 & $+$0.80 &  ...    & $+$0.51 & $+$0.87 & $+$0.77 & $-$1.15 & ...     & ...     \\
2M13435919$-$1807213 & $+$0.38 &    ...  & $+$0.77 & ...     & $+$0.28 & $+$0.82 & $+$0.83 & $-$1.00 &    ...  & $+$1.82 & $+$0.08 &    ...  & $+$0.49 &  ...    & $+$0.57 & $+$0.93 & $+$0.82 & $-$1.18 & ...     & $+$1.26 \\
2M17183459$+$4302520 & $+$0.15 & $+$0.56 & $+$0.79 & ...     & $+$0.37 & $+$1.03 & $+$1.03 & $-$1.05 & $+$1.12 &  ...    & $+$0.17 & $+$0.50 & $+$0.74 &  ...    & $+$0.61 & $+$1.25 & $+$1.09 & $-$1.13 & $+$0.95 & ...     \\
2M18091354$-$2810087 & $-$0.11 & $+$0.35 & $+$0.38 & $+$0.15 & $+$0.65 & $+$1.12 & $+$0.70 & $-$1.03 & $+$0.67 & $+$0.67 &    ...  &    ...  &    ...  &  ...    &    ...  &    ...  &    ...  & ...     & ...     & ...     \\
2M13303961$+$2719096 & $+$0.31 & $+$0.52 & $+$0.77 & ...     & $+$0.26 & $+$0.63 & $+$0.59 & $-$0.96 & $+$0.65 &  ...    & $+$0.23 & $+$0.56 & $+$0.75 &  ...    & $+$0.38 & $+$0.68 & $+$0.61 & $-$1.01 & $+$0.59 & ...     \\
2M15243300$+$2819313 & $+$0.09 & $+$0.19 & $+$0.37 & $+$0.23 & $+$0.33 & $+$0.75 & $+$0.68 & $-$0.91 & $+$0.37 &  ...    & $-$0.04 & $+$0.36 & $+$0.49 & $+$0.25 & $+$0.60 & $+$0.89 & $+$0.67 & $-$0.91 & $+$0.29 & ...     \\
2M16130340$-$3144580 & $+$0.35 & $+$0.46 & $+$0.67 & ...     & $+$0.33 & $+$0.77 & $+$0.70 & $-$0.94 & $+$1.27 &  ...    & $+$0.36 & $+$0.45 & $+$0.67 &  ...    & $+$0.39 & $+$0.80 & $+$0.66 & $-$0.95 & $+$1.06 & ...     \\
2M16483594$-$0150117 & $+$0.10 & $+$0.25 & $+$0.48 & ...     & $+$0.27 & $+$0.83 & $+$0.68 & $-$0.92 & $+$0.74 &  ...    & $-$0.01 & $+$0.35 & $+$0.52 &  ...    & $+$0.48 & $+$0.92 & $+$0.71 & $-$0.98 & $+$0.42 & ...     \\
2M17161376$-$2910175 & $+$0.04 & $+$0.10 & $+$0.33 & ...     & $+$0.28 & $+$0.70 & $+$0.62 & $-$0.91 & $+$0.70 &  ...    & $-$0.10 & $+$0.41 & $+$0.68 &  ...    & $+$0.58 & $+$0.95 & $+$0.48 & $-$0.90 & $+$0.56 & ...     \\
2M17171046$-$3007398 & $-$0.10 & $+$0.35 & $+$0.32 & ...     & $+$0.45 & $+$0.78 & $+$0.64 & $-$0.92 & $+$0.60 & $+$0.59 & $-$0.20 & $+$0.48 & $+$0.44 &  ...    & $+$0.65 & $+$0.92 & $+$0.60 & $-$0.99 & $+$0.32 & $+$0.62 \\
2M17421220$-$3443594 & $-$0.09 & $+$0.31 & $+$0.37 & ...     & $+$0.29 & $+$0.68 & $+$0.69 & $-$0.93 & $+$0.29 & $+$0.74 & $+$0.16 & $+$0.02 & $+$0.22 &  ...    & $+$0.32 & $+$0.42 & $+$0.70 & $-$1.27 & $-$0.01 & $+$0.50 \\
2M18024132$-$2940238 & $+$0.12 & $+$0.66 & $+$0.33 & ...     & $+$0.25 & $+$0.61 & $+$0.62 & $-$0.96 & $+$0.50 &  ...    & $-$0.03 & $+$0.71 & $+$0.29 &  ...    & $+$0.48 & $+$0.67 & $+$0.66 & $-$1.05 & $+$0.24 & ...     \\
2M16543450$-$0429397 & $+$0.05 &    ...  & $+$0.77 & ...     & $+$0.37 & $+$0.87 & $+$0.92 & $-$0.79 & $+$0.76 &  ...    & $+$0.16 &    ...  & $+$0.67 &  ...    & $+$0.43 & $+$0.90 & $+$0.98 & $-$0.89 & $+$0.46 & ...     \\
					\hline
				\end{tabular}  \label{Table2}
							\tablefoot{$^{*}$ denotes the elemental abundances determined by adopting spectroscopic atmospheric parameters ($T_{eff}^{ASPCAP}$ and $\log$ \textit{g$^{ASPCAP}$}), while $^{\dagger}$ refers to those chemical species determined by adopting photometric atmospheric parameters ($T_{eff}^{pho}$ and $\log$ \textit{g$^{iso}$}) as listed in Table \ref{Table1}, and in both cases the overall-metallicity ([M/H]) is used as first guess. The reference Solar abundances are from \citet{Asplund2005}, except for Ce II \citep{Cunha2017} and Nd II \citep{Hasselquist2016}, for which we have adopted the Solar abundances from \citet{Grevesse2015}. It is important to note, that [Na/Fe] value provided in this table corresponds to the determined upper limit to the abundance where the line intensity is comparable to the variance. We warning caution to use [Na/Fe].}
			\end{tiny}
		\end{sidewaystable*}

		\begin{table*}
			\begin{tiny}
				\begin{center}
					\setlength{\tabcolsep}{1.8mm}  
					\caption{Sensitivity to typical uncertainties in atmospheric parameters and standard deviation between lines of the same species for twenty six randomly selected stars in the \texttt{Jurassic} structure.}
					\begin{tabular}{ccccccccccc}
						\hline 
						\hline 
						APOGEE\_ID         & Fe I & $^{12}$C &  $^{14}$N   & $^{16}$O & Mg I & Al I & Si I & Na I & Ce II & Nd II \\
						\hline
						\hline
						2M09133506$+$2248579 &  0.15 &  ... &  ... & ... & 0.09 & 0.17 & 0.09 &  ... & ... & ...\\
						\hline
						$\sigma_{\rm [X/H],T_{eff}}$             &   0.04   & ...  & ...   & ...    & 0.03 & 0.05  & 0.02   &  ... & ...  &  ... \\
						$\sigma_{\rm [X/H], log \textit{g}}$  &   0.02  & ...   & ...   & ...   & 0.08  & 0.03  & 0.05  & ...   & ...  & ... \\
						$\sigma_{\rm [X/H], \xi_{t}}$ &    0.002   & ...     & ...    & ...    & 0.007  & 0.002 & 0.004  & ...    & ...   & ... \\
						$\sigma_{mean}$  &  0.14 &  ...  &  ...   &  ...   &  0.02 &  0.16  &   0.08  &  ...   &   ...   &  ... \\
						\hline
						2M12242950$+$4408525               &   0.21 & ... & ... & ... & 0.10 & 0.04 & 0.07 & ... & ... & ... \\
						\hline
						$\sigma_{\rm [X/H],T_{eff}}$             &   0.08   & ... & ...  & ...    & 0.06 & 0.03 & 0.04   &  ... & ... &  ... \\
						$\sigma_{\rm [X/H], log \textit{g}}$  &   0.01  & ... & ...   & ...   & 0.08  & 0.02  & 0.02  & ...   & ... & ... \\
						$\sigma_{\rm [X/H], \xi_{t}}$ &   0.001   & ...   & ...    & ...    & 0.002  & 0.001 & 0.001  & ...    & ...   & ...\\
						$\sigma_{mean}$  &   0.19 &  ...  &  ...   &  ...   &  0.03 &  0.01  &   0.05  &  ...   &   ...   &  ... \\
						\hline
						2M12443130$-$0900220              &   0.13 & 0.49 & ... & ... & 0.12 & 0.09 & 0.11 & ... & ... & ... \\
						\hline
						$\sigma_{\rm [X/H],T_{eff}}$              &  0.01  & 0.30  & ...   & ...  & 0.03  & 0.02  & 0.02 &  ... & ...  & ... \\
						$\sigma_{\rm [X/H], log \textit{g}}$  &  0.02  & 0.39  & ...   & ...  & 0.11  & 0.02  & 0.04  & ...  & ...  & ...\\
						$\sigma_{\rm [X/H], \xi_{t}}$ &  0.001   & 0.001   & ...    & ...  & 0.002  & 0.002 & 0.001  & ...    & ...   & ...\\
						$\sigma_{mean}$  &  0.13 &  ...  &  ...   &  ...  &  0.02 &  0.09  &   0.10  &  ...   &   ...   &  ... \\
						\hline
						2M13472354$+$2210562              &  0.16 & 0.09 & ... & ... & 0.13 & 0.14 & 0.13 & ... & 0.17 & ... \\
						\hline
						$\sigma_{\rm [X/H],T_{eff}}$             &   0.05  &    0.05  &  ...    &  ...    &    0.05  &     0.05 &     0.05  &     ...   & 0.01  &  ... \\
						$\sigma_{\rm [X/H], log \textit{g}}$  &   0.01  &    0.07  &   ...   &   ...  &   0.12  &    0.13  &     0.04  &     ... & 0.08     &   ... \\
						$\sigma_{\rm [X/H], \xi_{t}}$ &  0.002   & 0.003   & ...    & ...  & 0.005  & 0.005 & 0.004  & ...    & 0.155   & ...\\
						$\sigma_{mean}$  &   0.15 &  ...  &  ...   &  ...   &  0.02 &  ...   &   0.11  &  ...   &   ...   &  ... \\
						\hline
						2M13535604$+$4437076  &   0.09 & 0.48 & 0.39 & 0.37 & 0.13 & 0.15 & 0.14 & ... & 0.17 & ... \\
						\hline
						$\sigma_{\rm [X/H],T_{eff}}$             &   0.03   &    0.39   &    0.36   &    0.11        &    0.07 &     0.07 &     0.05  &     ...            &     0.02  &     ... \\
						$\sigma_{\rm [X/H], log \textit{g}}$  &   0.01  &    0.18    &    0.15   &    0.10        &   0.10  &    0.08  &     0.02  &     ...             &    0.12  & ... \\
						$\sigma_{\rm [X/H], \xi_{t}}$ &   0.003   & 0.001   & 0.001  & 0.001  & 0.004  & 0.005 & 0.010  & ...    & 0.003   & ... \\
						$\sigma_{mean}$  &  0.08 &  0.21 &  0.05  &  0.34  &  0.04 &  0.10  &   0.13  &  ...   &   0.12  &  ... \\
						\hline
						2M14533964$+$4506180              &   0.15 & ... & ... & ... & 0.12 & 0.22 & 0.07 & ... & 0.54 & ... \\
						\hline
						$\sigma_{\rm [X/H],T_{eff}}$              &  0.05  &    ...   & ...  & ... & 0.07    &  0.18  & 0.03  & ...    & 0.36 &  ... \\
						$\sigma_{\rm [X/H], log \textit{g}}$  &   0.01  &    ...   & ... & ...  & 0.06  & 0.12   & 0.01  & ...   & 0.09  & ... \\
						$\sigma_{\rm [X/H], \xi_{t}}$         &   0.001   & ...   & ...    & ...  & 0.004  & 0.004 & 0.004  & ...    & 0.395   & ... \\
						$\sigma_{mean}$                               &   0.14 &  ...  &  ...   &  ...   &  0.08 &  ...   &   0.06  &  ...   &   ...   &  ... \\
						\hline
						2M15170852$+$4033475                 &   0.18 & ... & ... & ... & 0.18 & 0.18 & 0.07 & ... & ... & ... \\
						\hline
						$\sigma_{\rm [X/H],T_{eff}}$             &   0.03     & ...  &  ...   & ...   & 0.07 &  0.06  & 0.03  & ...  &  ...   & ... \\
						$\sigma_{\rm [X/H], log \textit{g}}$  &  0.01     &  ...  & ...   & ...   & 0.09  & 0.04  & 0.03  & ...   &  ...   & ...\\
						$\sigma_{\rm [X/H], \xi_{t}}$ &  0.001   & ...     & ...    & ...    & 0.007  & 0.001 & 0.002  & ...    & ...     & ...\\
						$\sigma_{mean}$  &   0.18 &  ...  &  ...   &  ...   &  0.14 &  0.16  &   0.05  &  ...   &   ...   &  ... \\
						\hline
						2M15275895$+$4226412                &   0.09 & 0.08 & 0.16 & 0.16 & 0.13 & ... & 0.07 & ... & 0.13 & 0.13 \\
						\hline
						$\sigma_{\rm [X/H],T_{eff}}$             &  0.02  &    0.02  &    0.12  &    0.16  &    0.08  &    ...  &     0.01 &    ...       &     0.09  &     0.05\\
						$\sigma_{\rm [X/H], log \textit{g}}$  &  0.03  &   0.07  &    0.02   &    0.01  &    0.04  &   ...   &     0.04  &   ...        &      0.06  &    0.04\\
						$\sigma_{\rm [X/H], \xi_{t}}$ &   0.005   & 0.001   & 0.001  & 0.002  & 0.014  & ... & 0.009  & ...    & 0.008   & 0.001 \\
						$\sigma_{mean}$  &   0.08 &  0.04 &  0.11  &  0.03  &  0.10 &  ...   &   0.05  &  ...   &   0.08  &  0.11 \\
						\hline
						2M16013102$+$0618450                 &   0.12 & 0.04 & 0.35 & 0.22 & 0.11 & 0.14 & 0.07 & ... & 0.14 & ...\\
						\hline
						$\sigma_{\rm [X/H],T_{eff}}$             &   0.05  &    0.01 &    0.12  &    0.13  &    0.07  &     0.13  &     0.03  &    ...   &     0.07  &   ...\\
						$\sigma_{\rm [X/H], log \textit{g}}$  &   0.02  &   0.04  &    0.06  &    0.04  &   0.06  &      0.02  &     0.01  &     ...  &     0.08  & ... \\
						$\sigma_{\rm [X/H], \xi_{t}}$ &   0.021   & 0.001   & 0.155  & 0.002  & 0.008  & 0.004 & 0.014  & ...    & 0.066   & ... \\
						$\sigma_{mean}$  &   0.11 &  0.01 &  0.28  &  0.17  &  0.06 &  0.06  &   0.06  &  ...   &   0.07  &  ... \\
						\hline
						2M16441013$-$1850478                 &   0.10 & 0.17 & 0.14 & 0.17 & 0.15 & 0.12 & 0.08 & ... & 0.15 & 0.43 \\
						\hline
						$\sigma_{\rm [X/H],T_{eff}}$              &   0.03  &    0.01 &    0.09  &    0.11  &    0.07  &     0.07  &     0.01 &   ...           &     0.03   &     0.04\\
						$\sigma_{\rm [X/H], log \textit{g}}$   &   0.01  &   0.08  &    0.05  &    0.01  &    0.12  &     0.07  &     0.01  &  ...            &    0.11  &     0.43 \\
						$\sigma_{\rm [X/H], \xi_{t}}$ &   0.004   & 0.008   & 0.005  & 0.001  & 0.008  & 0.008 & 0.005  & ...    & 0.005   & 0.016\\
						$\sigma_{mean}$  &   0.10 &  0.15 &  0.09  &  0.13  &  0.06 &  0.07  &   0.08  &  ...   &   0.09  &  0.04 \\
						\hline
						2M16595910$+$1127496               &   0.12 & 0.09 & 0.29 & ... & 0.14 & 0.35 & 0.08 & ... & 0.28 & ...\\
						\hline
						$\sigma_{\rm [X/H],T_{eff}}$              &  0.04  &    0.07 &    0.15   &    ...   & 0.08 & 0.21  &    0.02  &   ...   & 0.04   &     ... \\
						$\sigma_{\rm [X/H], log \textit{g}}$  &   0.01  &   0.04  &    0.15  &     ...  & 0.10  & 0.21  &   0.03  &   ...   & 0.25   &     ... \\
						$\sigma_{\rm [X/H], \xi_{t}}$ &   0.004   & 0.004   & 0.003  & ...    & 0.006  & 0.005 & 0.004  & ...    & 0.001   & ... \\
						$\sigma_{mean}$  &   0.11 &  0.03 &  0.20  &  ...   &  0.06 &  0.19  &   0.07  &  ...   &   0.13  &  ... \\
						\hline
						2M17255079$-$2029099               &   0.09 & 0.22 & 0.17 & 0.17 & 0.16 & 0.11 & 0.07& ... & 0.15 & 0.08 \\
						\hline
						$\sigma_{\rm [X/H],T_{eff}}$             &  0.03   &    0.07  &    0.09 &    0.14 &    0.09 &     0.09  &     0.01 &   ...    &     0.04  &  0.03 \\
						$\sigma_{\rm [X/H], log \textit{g}}$  &   0.02  &    0.08  &   0.05  &   0.01  &   0.12  &    0.05  &      0.04  & ...     &     0.13  & 0.02 \\
						$\sigma_{\rm [X/H], \xi_{t}}$ &   0.007   & 0.008   & 0.006  & 0.004  & 0.010  & 0.011 & 0.007  & ...    & 0.005   & 0.001\\
						$\sigma_{mean}$  &   0.09 &  0.19 &  0.13  &  0.10  &  0.06 &  0.05  &   0.06  &  ...   &   0.06  &  0.07 \\
						\hline
						2M17265466$-$1331522               &   0.13 & 0.15 & 0.17 & 0.17 & 0.18 & 0.11 & 0.13 & ... & 0.19 & 0.19\\
						\hline
						$\sigma_{\rm [X/H],T_{eff}}$             &   0.03  & 0.01  &  0.09  &    0.14  & 0.08   & 0.07  & 0.01 &  ...   &   0.05  & 0.14  \\
						$\sigma_{\rm [X/H], log \textit{g}}$  &   0.02  & 0.08  & 0.03  &     0.01  & 0.14  & 0.08  & 0.02  & ...    &  0.17  & 0.08 \\
						$\sigma_{\rm [X/H], \xi_{t}}$ &   0.005   & 0.001   & 0.001  & 0.001  & 0.008  & 0.006 & 0.024  & ...    & 0.019   & 0.001 \\
						$\sigma_{mean}$  &   0.12 &  0.13 &  0.14  &  0.10  &  0.08 &  ...   &   0.13  &  ...   &   0.08  &  0.10 \\
						\hline
					\end{tabular}  \label{Table3}\\
				\end{center}
			\end{tiny}
		\end{table*}

		\begin{table*}
			\begin{tiny}
				\begin{center}

						\setlength{\tabcolsep}{1.8mm}  
					\caption{Continue.}
						\begin{tabular}{ccccccccccc}
						\hline
						2M17513049$+$5801309                &   0.13& 0.15 & 0.18 & 0.19 & 0.09 & 0.07 & 0.08 & ... & 0.06 & ... \\
						\hline
						$\sigma_{\rm [X/H],T_{eff}}$             &   0.04  &  0.03   & 0.16 &    0.11  &    0.06   & 0.04  & 0.03   &   ...  & 0.02  &     ... \\
						$\sigma_{\rm [X/H], log \textit{g}}$  &   0.01  & 0.02   & 0.06  &    0.03  &     0.05  & 0.03  & 0.01  &   ...  & 0.06  & ...\\
						$\sigma_{\rm [X/H], \xi_{t}}$ &   0.004   & 0.001   & 0.002  & ...    & 0.004  & 0.005 & 0.003  & ...  & 0.001   & ... \\
						$\sigma_{mean}$  &  0.12 &  0.15 &  0.05  &  0.16  &  0.04 &  0.05  &   0.07  &  ...   &   ...   &  ... \\
						\hline
						2M17572447$-$3056414              &  0.16 & 0.08 & ... & 0.16 & 0.17 & 0.12 & 0.09 & ... & 0.42 & ... \\
						\hline
						$\sigma_{\rm [X/H],T_{eff}}$              &  0.06  & 0.03 & ...   & 0.13  & 0.07 & 0.09   & 0.03  & ...  &  0.14  & ... \\
						$\sigma_{\rm [X/H], log \textit{g}}$  &  0.02  & 0.07  & ...  & 0.03  & 0.07  & 0.06  & 0.02  & ...   & 0.38  & ... \\
						$\sigma_{\rm [X/H], \xi_{t}}$ &   0.003   & 0.001   & ...    & 0.001  & 0.008  & 0.008 & 0.007  & ...    & 0.002   & ... \\
						$\sigma_{mean}$  &   0.15 &  0.03 &  ...   &  0.08  &  0.14 &  0.06  &   0.08  &  ...   &   0.10  &  ... \\
						\hline
						2M18013098$-$3307263  &  0.09 & 0.12  & 0.21 & 0.16 & 0.15 & 0.16 & 0.09 & 0.01 & 0.07 & 0.07 \\
						\hline
						$\sigma_{\rm [X/H],T_{eff}}$              &  0.04 & 0.01   & 0.14   &    0.15 &    0.11  &  0.15  & 0.06  & 0.01 &  0.01  & 0.02 \\
						$\sigma_{\rm [X/H], log \textit{g}}$  &  0.02  & 0.08  & 0.03   &    0.01  &    0.09  & 0.04  & 0.02  & 0.01  & 0.03  & 0.03 \\
						$\sigma_{\rm [X/H], \xi_{t}}$ &   0.007   & 0.001   & 0.012  & 0.003  & 0.020  & 0.018 & 0.016  & 0.004  & 0.011   & 0.055 \\
						$\sigma_{mean}$  &   0.08 &  0.09 &  0.15  &  0.06  &  0.04 &  0.04  &   0.06  &  ...   &   0.06  &  ... \\
						\hline
						2M18043255$-$4819138  &  0.11 & 0.16 & 0.21 & 0.17 & 0.13 & 0.12& 0.09 & ... & ... & ... \\
						\hline
						$\sigma_{\rm [X/H],T_{eff}}$             &  0.03  & 0.01 &  0.08  & 0.15  & 0.09 & 0.10  &  0.02  & ...  &  ...  & ... \\
						$\sigma_{\rm [X/H], log \textit{g}}$  & 0.03  & 0.09  & 0.04  & 0.01  & 0.04  & 0.02  & 0.07  & ...   & ...  &  ... \\
						$\sigma_{\rm [X/H], \xi_{t}}$ &   0.005   & 0.001   & 0.002  & 0.002  & 0.014  & 0.013 & 0.010  & ...    & ...   & ...\\
						$\sigma_{mean}$  &   0.10 &  0.13 &  0.19  &  0.07  &  0.08 &  0.07  &   0.06  &  ...   &   ...  &  ... \\
						\hline
						2M18151248$-$4403407  &   0.09 & 0.13 & 0.18 & 0.14 & 0.13 & 0.10 & 0.10 & 0.03 & ... & ... \\
						\hline
						$\sigma_{\rm [X/H],T_{eff}}$             &   0.02  &    0.01  &    0.11  &  0.13  & 0.08 &  0.08 &  0.01 &  0.03  &   ...  &    ...\\
						$\sigma_{\rm [X/H], log \textit{g}}$  &   0.02  &    0.08  &    0.04  & 0.01  & 0.10  & 0.04  & 0.01  & 0.01  &   ...  & ... \\
						$\sigma_{\rm [X/H], \xi_{t}}$ &   0.006   & 0.001   & 0.003  & 0.001  & 0.013  & 0.011 & 0.011  & 0.001  & ...     & ... \\
						$\sigma_{mean}$  &   0.09 &  0.10 &  0.14  &  0.05  &  0.04 &  0.05  &   0.10  &  ...   &   ...   &  ... \\
						\hline
						2M18453994$-$3010465  &   0.19 & 0.09 & 0.13 & 0.16 & 0.13 & 0.12 & 0.15 & ... & 0.09 & 0.26 \\
						\hline
						$\sigma_{\rm [X/H],T_{eff}}$              &   0.03  & 0.05  & 0.05 &  0.14 &  0.04 &  0.09 &  0.03  & ... &  0.04  & 0.02 \\
						$\sigma_{\rm [X/H], log \textit{g}}$  &   0.01  & 0.07  & 0.03  & 0.02  & 0.08  & 0.07  & 0.01  & ...  & 0.06  & 0.25\\
						$\sigma_{\rm [X/H], \xi_{t}}$ &   0.005   & 0.001   & 0.015  & 0.003  & 0.016  & 0.013 & 0.008  & ...  & 0.003   & 0.003 \\
						$\sigma_{mean}$  &   0.19 &  0.04 &  0.12  &  0.07  &  0.09 &  0.03  &   0.15  &  ...   &   0.05  &  0.08  \\
						\hline
						2M19105369$+$2717150  &   0.09 & 0.16 & 0.33 & 0.12 & 0.17 & 0.14 & 0.12 & ... & 0.27 & ... \\
						\hline
						$\sigma_{\rm [X/H],T_{eff}}$              &   0.03  & 0.03  & 0.25  & 0.07  & 0.07 &  0.05 &  0.09 & ...   & 0.02  &  ...\\
						$\sigma_{\rm [X/H], log \textit{g}}$  &   0.01  & 0.07  & 0.04  & 0.05  & 0.14  & 0.08  & 0.04  & ...   & 0.22  & ...\\
						$\sigma_{\rm [X/H], \xi_{t}}$ &   0.002   & 0.003   & 0.022  & 0.037  & 0.024  & 0.004 & 0.009  & ...    & 0.011   & ... \\
						$\sigma_{mean}$  &   0.08 &  0.14 &  0.21  &  0.08  &  0.05 &  0.10  &   0.06  &  ...   &   0.16  &  ... \\
						\hline
						2M19193412$-$2931210  &  0.13 & 0.21 & 0.24 & 0.23 & 0.09 & 0.07 & 0.08 & ... & 0.16 & ... \\
						\hline
						$\sigma_{\rm [X/H],T_{eff}}$              &  0.04  &  0.03  &    0.16  &    0.12  &    0.06  &     0.05  &  0.01  & ...   & 0.05  &  ... \\
						$\sigma_{\rm [X/H], log \textit{g}}$  &   0.01  & 0.09  &     0.07  &    0.01  &    0.05  &     0.03  & 0.03  & ...   & 0.10  &   ... \\
						$\sigma_{\rm [X/H], \xi_{t}}$ &   0.001   & 0.031   & 0.026  & 0.003  & 0.014  & 0.008 & 0.005  & ...    & 0.002   & ... \\
						$\sigma_{mean}$  &  0.12 &  0.19 &  0.16  &  0.20  &  0.06 &  0.04  &   0.07  &  ...   &   0.11  &  ... \\
						\hline
						2M19214936$-$1232462  &   0.09 & 0.09 & 0.17 & 0.19 & 0.12 & 0.09 & 0.12 & 0.04 & 0.17 & 0.21\\
						\hline
						$\sigma_{\rm [X/H],T_{eff}}$             &   0.04 &    0.01  & 0.14   & 0.18  & 0.11 &  0.08  & 0.03 & 0.03   & 0.05  & 0.04 \\
						$\sigma_{\rm [X/H], log \textit{g}}$  &  0.04  &    0.07  & 0.03  & 0.01  & 0.02  & 0.02  & 0.06  & 0.01  & 0.15  & 0.18 \\
						$\sigma_{\rm [X/H], \xi_{t}}$ &   0.008   & 0.004   & 0.006  & 0.004  & 0.019  & 0.018 & 0.005  & 0.002  & 0.008   & 0.006 \\
						$\sigma_{mean}$  &   0.07 &  0.07 &  0.09  &  0.05  &  0.05 &  0.05  &   0.10  &  0.02  &   0.05  &  0.09 \\
						\hline
						2M22045404$-$1148287  &  0.11 & 0.20 & 0.26 & 0.10 & 0.11 & 0.09 & 0.08 & ... & 0.13 & ... \\
						\hline
						$\sigma_{\rm [X/H],T_{eff}}$             &   0.03  &    0.05  &    0.15  &    0.08  &    0.07  &    0.07  &  0.01  & ...   &  0.03  & ... \\
						$\sigma_{\rm [X/H], log \textit{g}}$  &   0.01  &    0.11  &    0.20  &    0.01  &    0.08  &    0.03  & 0.02  &  ...   & 0.12  & ... \\
						$\sigma_{\rm [X/H], \xi_{t}}$ &   0.003   & 0.019   & 0.080  & 0.017  & 0.010  & 0.008 & 0.006  & ...    & 0.009   & ... \\
						$\sigma_{mean}$  &   0.11 &  0.16 &  0.03  &  0.06  &  0.03 &  0.06  &   0.08  &  ...   &   0.05  &  ... \\
						\hline
						2M16241820$-$2145485  &   0.14 & 0.10 & 0.18 & 0.19 & 0.18 & 0.27 & 0.16 & ... & 0.12 & 0.19 \\
						\hline
						$\sigma_{\rm [X/H],T_{eff}}$             &   0.07  &    0.01  &    0.15  &  0.19  & 0.15 &  0.23 &  0.06  & ...   & 0.07  & 0.11 \\
						$\sigma_{\rm [X/H], log \textit{g}}$  &  0.06   &    0.09  &    0.03  & 0.03  & 0.10  & 0.14  & 0.08  & ...   & 0.08  & 0.15 \\
						$\sigma_{\rm [X/H], \xi_{t}}$ &   0.002   & 0.006   & 0.015  & 0.004  & 0.017  & 0.016 & 0.009  & ...    & 0.005   & 0.006 \\
						$\sigma_{mean}$  &  0.11 &  0.05 &  0.09  &  0.05  &  0.02 &  ...   &   0.12  &  ...   &   0.05  &  0.05 \\
						\hline
						2M17295157$-$3737045  &  0.12 & 0.12 & 0.09 & 0.22 & 0.13 & 0.11 & 0.08 & ... & 0.19 & 0.13 \\
						\hline
						$\sigma_{\rm [X/H],T_{eff}}$             &    0.04  & 0.04 &  0.06 &  0.13  &  0.08 & 0.09  &  0.01  & ...   & 0.03  & 0.03 \\
						$\sigma_{\rm [X/H], log \textit{g}}$  &   0.02  & 0.05  & 0.03  & 0.03  &  0.05  & 0.02  & 0.02  & ...   & 0.15  & 0.12\\
						$\sigma_{\rm [X/H], \xi_{t}}$ &   0.003   & 0.001   & 0.001  & 0.001  & 0.012  & 0.011 & 0.005  & ...    & 0.002   & 0.001 \\
						$\sigma_{mean}$  &   0.11 &  0.10 &  0.07  &  0.17  &  0.09 &  0.06  &   0.08  &  ...   &   0.12  &  0.03 \\
						\hline
						2M02175837$-$7313144  &   0.12 & 0.09 & 0.21 & 0.18 & 0.16 & 0.14 & 0.19 & 0.10 & 0.24 & \\
						\hline
						$\sigma_{\rm [X/H],T_{eff}}$              &   0.03  & 0.01  & 0.17  &    0.17   &    0.08   & 0.13  &     0.11   &     0.10   & 0.17   &    ... \\
						$\sigma_{\rm [X/H], log \textit{g}}$  &   0.04  & 0.09  & 0.05   &    0.01   &    0.06  & 0.03   &     0.13   &     0.01  & 0.17    &    ... \\
						$\sigma_{\rm [X/H], \xi_{t}}$ &   0.007   & 0.001   & 0.003  & 0.004  & 0.014  & 0.013 & 0.014  & 0.011    & 0.010   & ... \\
						$\sigma_{mean}$  &  0.11 &  0.04 &  0.12  &  0.06  &  0.12 &  0.05  &   0.10  &  ...   &   0.05  &  ... \\
						\hline
						\hline
					\end{tabular}  \label{Table4}\\
				\end{center}
			\end{tiny}
		\end{table*}
		
				\begin{sidewaystable*}
			\begin{tiny}
				\setlength{\tabcolsep}{3.5mm}  
				\caption{Orbital elements of Si-rich stars.}
				\centering
				\begin{tabular}{|c|ccccccc|}
					\hline
					APOGEE\_ID           &        $Z_{max}$      &         $r_{peri}$       &  $r_{apo}$             &   eccentricity &   $L_{z}^{min}$                &    $L_{z}^{max}$                & Orbital sense                \\
					&        [kpc]    &        [kpc]      &  [kpc]            &    &   [km s kpc]                &    [km s kpc]                &              \\
					\hline
					\hline
					2M17513049$+$5801309 &  2.8  $\pm$ 0.1 (0.08) &  5.6$\pm$0.1 (0.02) &  9.1 $\pm$  0.1 (0.18) & 0.23$\pm$ 0.01 (0.01) & $-$166.0 $\pm$  2.0 (  3.7) &  $-$157.0 $\pm$  4.0 (1.6)  & Prograde             \\
					2M22184840$-$0431434 &  1.8  $\pm$ 0.2 (0.05) &  4.8$\pm$0.2 (0.18) &  8.2 $\pm$  0.1 (0.25) & 0.25$\pm$ 0.02 (0.01) & $-$149.0 $\pm$  6.0 (  2.3) &  $-$139.0 $\pm$  7.0 (5.7)  & Prograde             \\
					2M04463289$-$7336289 &  2.0  $\pm$ 0.2 (0.06) &  4.8$\pm$0.1 (0.04) &  8.4 $\pm$  0.1 (0.46) & 0.27$\pm$ 0.01 (0.02) & $-$149.0 $\pm$  2.0 ( 11.1) &  $-$141.0 $\pm$  1.5 (3.2)  & Prograde             \\
					2M12443130$-$0900220 &  1.9  $\pm$ 0.1 (0.03) &  4.2$\pm$0.1 (0.14) &  9.2 $\pm$  0.1 (0.35) & 0.36$\pm$ 0.01 (0.02) & $-$144.0 $\pm$ 1.0  (  7.5) &  $-$136.0 $\pm$  1.0 (4.2)  & Prograde             \\
					2M02175837$-$7313144 &  3.8  $\pm$ 0.4 (0.21) &  4.2$\pm$0.2 (0.25) &  9.2 $\pm$  0.2 (0.71) & 0.37$\pm$ 0.03 (0.03) & $-$134.0 $\pm$  4.5 (  6.9) &  $-$126.0 $\pm$  4.5 (8.0)  & Prograde             \\
					2M22045404$-$1148287 &  5.3  $\pm$ 0.1 (0.04) &  3.5$\pm$0.1 (0.16) &  7.9 $\pm$  0.1 (0.16) & 0.38$\pm$ 0.01 (0.01) & $-$105.0 $\pm$  3.0 (  0.8) &   $-$96.0 $\pm$  4.0 (4.6)  & Prograde             \\
					2M16595910$+$1127496 &  3.8  $\pm$ 1.5 (0.03) &  2.5$\pm$0.7 (0.14) &  6.5 $\pm$  0.9 (0.21) & 0.40$\pm$ 0.10 (0.02) &  $-$91.0 $\pm$ 16.5 (  1.8) &   $-$61.0 $\pm$ 21.1 (4.4)  & Prograde             \\
					2M15160402$+$4728327 &  1.7  $\pm$ 0.3 (0.10) &  3.5$\pm$0.1 (0.12) &  8.8 $\pm$  0.1 (0.26) & 0.43$\pm$ 0.01 (0.01) & $-$129.0 $\pm$  3.0 (  0.9) &  $-$113.0 $\pm$  4.0 (2.9)  & Prograde             \\
					2M19105369$+$2717150 &  2.0  $\pm$ 0.6 (0.04) &  4.5$\pm$0.1 (0.01) & 12.3 $\pm$  2.4 (0.10) & 0.45$\pm$ 0.06 (0.01) & $-$169.0 $\pm$ 14.5 (  0.4) &  $-$160.0 $\pm$ 18.0 (1.2)  & Prograde             \\
					2M13303961$+$2719096 &  3.9  $\pm$ 0.3 (0.18) &  3.3$\pm$0.3 (0.19) &  9.5 $\pm$  0.2 (0.33) & 0.47$\pm$ 0.04 (0.03) & $-$119.0 $\pm$  6.6 (  5.4) &  $-$107.0 $\pm$  9.0 (4.3)  & Prograde             \\
					2M16005847$-$3708334 &  2.3  $\pm$ 0.4 (0.03) &  1.4$\pm$1.4 (0.05) &  4.4 $\pm$  1.2 (0.15) & 0.51$\pm$ 0.25 (0.01) &  $-$68.0 $\pm$ 37.0 (  5.3) &   $-$35.0 $\pm$ 33.5 (0.8)  & Prograde             \\
					2M16543450$-$0429397 &  4.0  $\pm$ 1.0 (0.06) &  1.6$\pm$0.8 (0.27) &  5.6 $\pm$  0.5 (0.16) & 0.54$\pm$ 0.19 (0.07) &  $-$67.0 $\pm$ 16.5 (  1.6) &   $-$36.0 $\pm$ 19.0 (5.4)  & Prograde             \\
					2M18453994$-$3010465 &  2.9  $\pm$ 0.2 (0.06) &  1.3$\pm$0.8 (0.24) &  4.7 $\pm$  0.9 (0.21) & 0.56$\pm$ 0.20 (0.06) &  $-$51.0 $\pm$ 20.0 (  4.4) &   $-$26.0 $\pm$ 26.0 (6.1)  & Prograde             \\
					2M17183459$+$4302520 &  7.0  $\pm$ 3.7 (0.42) &  2.2$\pm$1.3 (0.16) &  9.9 $\pm$  1.9 (0.66) & 0.62$\pm$ 0.09 (0.00) &  $-$90.0 $\pm$ 26.0 (  2.8) &   $-$74.0 $\pm$ 35.5 (2.3)  & Prograde             \\
					2M15243300$+$2819313 &  5.9  $\pm$ 0.4 (0.08) &  1.4$\pm$0.5 (0.17) &  8.0 $\pm$  0.2 (0.40) & 0.69$\pm$ 0.10 (0.02) &  $-$61.0 $\pm$ 17.5 (  8.4) &   $-$38.0 $\pm$ 16.5 (5.8)  & Prograde             \\
					2M22375002$-$1654304 &  5.0  $\pm$ 2.7 (0.04) &  1.1$\pm$1.2 (0.14) &  8.2 $\pm$  0.5 (0.26) & 0.74$\pm$ 0.18 (0.01) &  $-$57.0 $\pm$ 55.1 (  6.9) &   $-$35.0 $\pm$ 55.0 (6.1)  & Prograde             \\
					2M17265466$-$1331522 &  2.6  $\pm$ 0.7 (0.08) &  0.4$\pm$1.3 (0.14) &  4.2 $\pm$  2.3 (0.02) & 0.78$\pm$ 0.27 (0.03) &  $-$46.0 $\pm$ 49.5 (  7.2) &   $-$12.0 $\pm$ 43.5 (4.3)  & Prograde             \\
					2M16483594$-$0150117 &  2.6  $\pm$ 0.8 (0.31) &  0.6$\pm$0.7 (0.03) &  5.3 $\pm$  0.6 (0.05) & 0.79$\pm$ 0.16 (0.00) &  $-$50.0 $\pm$ 28.0 (  2.0) &   $-$20.0 $\pm$ 24.5 (0.8)  & Prograde             \\
					2M18043255$-$4819138 &  2.2  $\pm$ 0.5 (0.02) &  0.4$\pm$0.7 (0.20) &  4.0 $\pm$  0.6 (0.06) & 0.80$\pm$ 0.20 (0.07) &  $-$54.0 $\pm$ 21.5 (  3.0) &   $-$10.0 $\pm$ 18.5 (4.5)  & Prograde             \\
					2M17171046$-$3007398 &  0.7  $\pm$ 0.3 (0.03) &  0.1$\pm$0.1 (0.06) &  2.0 $\pm$  0.9 (0.05) & 0.81$\pm$ 0.11 (0.05) &  $-$26.0 $\pm$ 14.5 (  0.4) &    $-$3.0 $\pm$  4.5 (1.2)  & Prograde             \\
					2M19193412$-$2931210 &  2.8  $\pm$ 0.7 (0.05) &  0.3$\pm$0.6 (0.05) &  4.3 $\pm$  1.0 (0.29) & 0.83$\pm$ 0.18 (0.02) &  $-$38.0 $\pm$ 22.1 (  2.4) &    $-$6.0 $\pm$ 16.5 (1.2)  & Prograde             \\
					2M13590229$-$0831592 &  6.4  $\pm$ 4.0 (0.13) &  0.7$\pm$1.1 (0.04) &  8.4 $\pm$  1.7 (0.31) & 0.86$\pm$ 0.20 (0.01) &  $-$19.0 $\pm$ 67.5 (  1.6) &    $-$2.0 $\pm$ 52.0 (2.1)  & Prograde             \\
					2M17295157$-$3737045 & 45.6  $\pm$42.5 (0.09) &  5.3$\pm$2.1 (0.01) & 82.5 $\pm$ 68.7 (0.17) & 0.87$\pm$ 0.05 (3.42) & $-$262.0 $\pm$ 87.1 (  0.4) &  $-$261.0 $\pm$ 88.1 (0.4)  & Prograde             \\
					2M18220651$-$3835466 &  1.4  $\pm$ 0.2 (0.01) &  0.1$\pm$0.4 (0.09) &  2.9 $\pm$  0.9 (0.08) & 0.88$\pm$ 0.16 (0.05) &  $-$37.0 $\pm$ 30.0 (  1.6) &    $-$3.0 $\pm$ 18.5 (2.4)  & Prograde             \\
					2M18091354$-$2810087 &  1.5  $\pm$ 0.4 (0.03) &  0.1$\pm$0.2 (0.01) &  2.1 $\pm$  0.9 (0.23) & 0.90$\pm$ 0.15 (0.01) &  $-$23.0 $\pm$ 10.1 (  2.3) &    $-$1.0 $\pm$  6.1 (0.4)  & Prograde             \\
					2M18013098$-$3307263 &  1.0  $\pm$ 0.2 (0.01) &  0.0$\pm$0.1 (0.02) &  2.0 $\pm$  0.9 (0.02) & 0.90$\pm$ 0.06 (0.02) &  $-$24.0 $\pm$ 11.5 (  0.8) &    $-$1.0 $\pm$  4.1 (0.8)  & Prograde             \\
					2M13472354$+$2210562 &  3.6  $\pm$ 0.7 (0.85) &  0.2$\pm$0.3 (0.01) &  9.3 $\pm$  0.1 (0.09) & 0.95$\pm$ 0.07 (0.01) &  $-$31.0 $\pm$ 20.5 (  5.8) &    $-$8.0 $\pm$ 18.0 (2.1)  & Prograde             \\
					2M13535604$+$4437076 &  1.6  $\pm$ 0.1 (0.03) &  3.1$\pm$0.6 (0.01) &  8.8 $\pm$  0.1 (0.01) & 0.47$\pm$ 0.06 (0.01) &    111.0 $\pm$ 15.5 (  0.1) &     115.0 $\pm$ 15.0 (0.4)  & Retrograde           \\
					2M14533964$+$4506180 &  3.0  $\pm$ 0.6 (0.19) &  2.2$\pm$0.6 (0.05) &  8.5 $\pm$  0.1 (0.10) & 0.59$\pm$ 0.09 (0.01) &     79.0 $\pm$ 16.0 (  2.6) &      91.0 $\pm$ 17.0 (0.4)  & Retrograde           \\
					2M15170852$+$4033475 &  2.3  $\pm$ 0.5 (0.21) &  1.8$\pm$0.6 (0.01) &  8.1 $\pm$  0.2 (0.04) & 0.62$\pm$ 0.10 (0.01) &     71.0 $\pm$ 20.0 (  1.4) &      79.0 $\pm$ 16.0 (0.4)  & Retrograde           \\
					2M15275895$+$4226412 &  6.6  $\pm$ 2.0 (0.47) &  2.3$\pm$0.3 (0.02) & 12.6 $\pm$  2.5 (0.13) & 0.68$\pm$ 0.03 (0.01) &     94.0 $\pm$ 17.0 (  0.4) &     101.0 $\pm$ 14.0 (0.4)  & Retrograde           \\
					2M09133506$+$2248579 &  6.1  $\pm$ 2.6 (0.34) &  1.7$\pm$2.2 (0.03) & 15.3 $\pm$  1.8 (0.24) & 0.78$\pm$ 0.18 (0.01) &     73.0 $\pm$ 90.0 (  2.8) &      86.0 $\pm$ 86.6 (3.3)  & Retrograde           \\
					2M12242950$+$4408525 &  7.5  $\pm$ 0.8 (0.61) &  1.1$\pm$0.4 (0.10) & 11.0 $\pm$  1.0 (0.27) & 0.81$\pm$ 0.07 (0.01) &     43.0 $\pm$ 15.5 (  4.0) &      61.0 $\pm$  8.5 (4.3)  & Retrograde           \\
					2M17073023$-$2717147 &  4.4  $\pm$ 6.3 (0.40) &  0.7$\pm$0.4 (0.05) & 11.2 $\pm$  5.9 (0.31) & 0.89$\pm$ 0.09 (0.01) &     24.0 $\pm$ 28.1 (  2.6) &      40.0 $\pm$ 26.0 (2.4)  & Retrograde           \\
					2M14513934$-$0602148 &  6.0  $\pm$ 1.6 (0.10) &  0.7$\pm$0.5 (0.11) &  6.5 $\pm$  0.9 (0.16) & 0.80$\pm$ 0.13 (0.02) &      0.1 $\pm$ 39.0 (  3.2) &      14.0 $\pm$ 33.5 (3.7)  & Retrograde           \\
					2M22480199$+$1411329 &  3.6  $\pm$ 0.5 (0.26) &  1.9$\pm$0.5 (0.10) &  8.7 $\pm$  0.2 (0.19) & 0.63$\pm$ 0.09 (0.02) &     70.0 $\pm$ 17.5 (  1.6) &      75.0 $\pm$ 15.5 (2.8)  & Retrograde           \\
					2M13435919$-$1807213 &  5.1  $\pm$ 0.9 (0.25) &  1.4$\pm$0.9 (0.05) &  7.8 $\pm$  0.3 (0.04) & 0.69$\pm$ 0.16 (0.01) &     47.0 $\pm$ 26.6 (  1.4) &      64.5 $\pm$ 20.6 (1.4)  & Retrograde           \\
					2M13315603$+$4354469 &  5.3  $\pm$ 0.7 (0.06) &  2.9$\pm$2.4 (0.05) & 10.0 $\pm$  0.6 (0.06) & 0.54$\pm$ 0.24 (0.01) &     99.0 $\pm$ 63.0 (  1.6) &     105.0 $\pm$ 55.5 (0.8)  & Retrograde           \\
					2M16013102$+$0618450 &  5.0  $\pm$ 0.8 (0.06) &  0.1$\pm$0.3 (0.09) &  6.1 $\pm$  0.5 (0.10) & 0.96$\pm$ 0.09 (0.02) &  $-$19.0 $\pm$ 20.0 (  7.1) &       1.0 $\pm$ 18.5 (5.3)  & P-R  \\
					2M16441013$-$1850478 &  3.1  $\pm$ 0.8 (0.23) &  0.1$\pm$0.1 (0.03) &  3.2 $\pm$  1.5 (0.20) & 0.95$\pm$ 0.08 (0.01) &   $-$2.0 $\pm$ 12.0 (  1.2) &      17.5 $\pm$ 10.5 (1.3)  & P-R  \\
					2M17255079$-$2029099 &  1.5  $\pm$ 0.1 (0.03) &  0.1$\pm$0.1 (8.86) &  1.7 $\pm$  0.5 (0.11) & 0.97$\pm$ 0.02 (7.85) &  $-$14.0 $\pm$  8.0 (  0.4) &       9.0 $\pm$  5.5 (2.0)  & P-R  \\
					2M17572447$-$3056414 &  1.0  $\pm$ 0.6 (0.03) &  0.1$\pm$0.2 (0.01) &  1.9 $\pm$  1.3 (0.00) & 0.96$\pm$ 0.11 (0.01) &   $-$9.0 $\pm$ 24.0 (  1.2) &       5.0 $\pm$ 17.0 (1.6)  & P-R  \\
					2M18151248$-$4403407 &  2.0  $\pm$ 0.4 (0.02) &  0.1$\pm$0.1 (0.01) &  2.7 $\pm$  0.8 (0.10) & 0.97$\pm$ 0.05 (0.01) &  $-$18.0 $\pm$ 20.1 (  4.7) &      11.0 $\pm$ 12.5 (4.1)  & P-R  \\
					2M19214936$-$1232462 &  4.7  $\pm$ 0.7 (0.13) &  0.4$\pm$0.5 (0.02) &  5.2 $\pm$  0.5 (0.11) & 0.83$\pm$ 0.15 (0.01) &   $-$7.0 $\pm$ 29.5 (  2.8) &      16.0 $\pm$ 23.0 (2.4)  & P-R  \\
					2M16241820$-$2145485 &  2.8  $\pm$ 0.1 (0.01) &  0.1$\pm$0.1 (0.01) &  3.2 $\pm$  0.7 (0.10) & 0.97$\pm$ 0.06 (0.01) &  $-$16.0 $\pm$ 23.0 (  2.3) &       7.0 $\pm$  9.0 (2.1)  & P-R  \\
					2M17171752$-$2148372 &  2.0  $\pm$ 0.5 (0.12) &  0.1$\pm$0.2 (0.06) &  2.7 $\pm$  1.0 (0.03) & 0.89$\pm$ 0.18 (0.03) &   $-$1.0 $\pm$ 15.5 (  2.9) &      25.0 $\pm$ 13.5 (1.6)  & P-R  \\
					2M18242843$-$3942097 &  2.1  $\pm$ 0.5 (0.01) &  0.1$\pm$0.1 (0.01) &  3.2 $\pm$  0.7 (0.01) & 0.97$\pm$ 0.08 (0.01) &  $-$22.0 $\pm$ 15.5 (  0.9) &       6.0 $\pm$ 14.0 (1.6)  & P-R  \\
					2M18280709$-$3900094 &  1.9  $\pm$ 0.1 (0.01) &  0.1$\pm$0.1 (0.01) &  2.5 $\pm$  1.0 (0.17) & 0.97$\pm$ 0.06 (0.01) &  $-$12.0 $\pm$ 20.1 (  3.2) &      13.0 $\pm$ 15.0 (5.1)  & P-R  \\
					2M16544476$-$3939140 &  1.3  $\pm$ 0.5 (0.09) &  0.1$\pm$0.6 (0.01) &  3.1 $\pm$  1.4 (0.01) & 0.96$\pm$ 0.18 (0.01) &  $-$20.0 $\pm$ 34.5 (  4.0) &       9.0 $\pm$ 28.5 (3.3)  & P-R  \\
					2M16130340$-$3144580 &  2.1  $\pm$ 0.6 (0.17) &  0.1$\pm$0.3 (0.01) &  3.9 $\pm$  0.5 (0.10) & 0.96$\pm$ 0.11 (0.01) &  $-$28.0 $\pm$ 24.5 (  1.4) &       1.0 $\pm$ 16.5 (0.8)  & P-R  \\
					2M17421220$-$3443594 &  3.6  $\pm$ 4.7 (0.05) &  0.4$\pm$0.8 (0.03) &  4.7 $\pm$  6.3 (0.23) & 0.85$\pm$ 0.12 (0.01) &  $-$28.0 $\pm$ 40.0 (  2.2) &       0.5 $\pm$ 43.5 (1.3)  & P-R  \\
					2M18024132$-$2940238 &  1.6  $\pm$ 0.7 (0.17) &  0.1$\pm$0.2 (0.04) &  2.3 $\pm$  1.5 (0.01) & 0.95$\pm$ 0.12 (0.02) &   $-$6.0 $\pm$ 16.0 (  1.2) &      18.0 $\pm$ 17.5 (0.8)  & P-R  \\
					\hline
				\end{tabular}  \label{Table5}
			\end{tiny}
		\end{sidewaystable*}

\end{appendix}

\end{document}